\documentclass{aa}

\usepackage{xcolor}
\usepackage[colorlinks=true,linkcolor=blue,citecolor=blue]{hyperref}
\usepackage{longtable}
\usepackage{multirow}
\usepackage{subcaption}
\usepackage{lscape}            

\usepackage{graphicx}
\usepackage{color}
\usepackage{txfonts}

\newcommand{\teff}{$T_\mathrm{eff}$}
\newcommand{\logg}{$\log g$}
\newcommand{\logage}{$\log(age)$}
\newcommand{\siglogage}{$\sigma_{\log(age)}$}
\newcommand{\bprp}{$G_\mathrm{BP}-G_\mathrm{RP}$}

\begin{document}

\title{Refining open cluster parameters with Gaia XP metallicities
 }

\author{M. Nizovkina  \inst{1}
\and S. S. Larsen \inst{1}
\and A. G. A. Brown \inst{2}
\and A. Helmi \inst{3}
}

\institute{Department of Astrophysics/IMAPP, Radboud University, PO Box 9010, 6500 GL Nĳmegen, The Netherlands\\
\email{m.nizovkina@astro.ru.nl}
\and Leiden Observatory, Leiden University, Einsteinweg 55, 2333 CC Leiden, The Netherlands
\and Kapteyn Astronomical Institute, University of Groningen, Landleven 12, 9747 AD Groningen, The Netherlands}

\date{Received 9 May 2025 / Accepted 14 September 2025}

\authorrunning{M. Nizovkina et al.}
\titlerunning{Refining open cluster parameters with Gaia XP metallicities}

\abstract
{Open clusters (OCs) are crucial objects for studying stellar evolution and Galactic dynamics due to the shared origin and composition of the stars within each cluster. The precision of cluster parameter determination has significantly improved with the availability of homogeneous photometric Gaia data; however, challenges such as age-metallicity degeneracy and lack of spectroscopic observations remain.
}
{We investigate whether metallicities derived from low-resolution Gaia XP spectra can be effectively used to break degeneracies and improve the accuracy of OC parameter determinations.
}
{We analysed 20 OCs using isochrone fitting methods on Gaia DR3 photometry and metallicity estimates from several Gaia XP-based catalogues. We used the synthetic cluster method, implemented through ASteCA, to derive age, distance modulus, and extinction using the approximate Bayesian computation (ABC).
}
{We compared the parameter estimates of these open clusters to the values obtained in other works through isochrone fitting with spectroscopically constrained metallicities or through neural network techniques applied only to the photometry. We found the systematic difference between Gaia XP derived metallicities and those obtained from high-resolution spectroscopy to be 0.1-0.15 dex. We found a systematic age difference of $<0.03 \pm 0.13$ dex compared to isochrone fitting using high-resolution spectroscopy, and $<0.08 \pm 0.21$ dex compared to neural network-based methods, and a median individual error, \siglogage, of $\sim$0.065 dex.
}
{Metallicities from Gaia XP spectra prove highly effective in constraining the parameters of open clusters that do not have a populated red giant branch (RGB), despite the low spectral resolution. When used with stringent quality cuts and incorporated as priors, they make it possible to determine ages comparable in precision to those based on high-resolution spectroscopy. Moreover, compared to studies of open clusters that use neural networks without metallicity constraints, our method yields smaller uncertainties. These results highlight the potential of Gaia data for accurate cluster parameter analysis and detailed Galactic studies, even in the absence of traditional spectroscopic observations.}


\maketitle

\section{Introduction}\label{sect:intro}
Open clusters (OC) are groups of stars that are located in the Galactic disc and have a common origin, therefore the same age and compositions, which makes them invaluable laboratories for studying stellar evolution and the dynamics of the galaxy. They used to be identified from the ground-based data \citep{dias2002, kharchenko2005}, which was limited to brighter, closer and more populated clusters. 

The open cluster census was tremendously improved by data from the Gaia satellite \citep{gaia}, as it provided homogeneous astrometric and photometric data of unprecedentedly high precision for the whole sky. This led to a significant increase in the amount of identified OCs based on their precise positions, proper motions and parallaxes using clustering methods, starting from $\sim$3000 OCs known pre-Gaia and more than 7000 known afterwards \citep[e.g.][]{sim2019, liu2019, monteiro2019}. A large portion of the newly discovered clusters was caused by the development of machine learning techniques that were applied to OCs to perform ‘blind searches’ \citep{castro2019}. Most notably, \citet{cantat2018} compiled a homogeneous catalogue of 1229 OCs by applying an unsupervised membership determination algorithm to Gaia Data Release 2 \citep{gaia2018}. Their method included using a clustering method DBSCAN on astrometric data and applying a neural network to the colour-magnitude diagram (CMD) of clusters to validate the membership. \citet{hunt2021} took this work further first by comparing various clustering methods and establishing that instead HDBSCAN shows the most sensitivity to OCs in Gaia data. In their follow-up work \citep[hereafter HR23]{hunt2023} they applied HDBSCAN together with the neural network to Gaia DR3 data, thus compiling a comprehensive catalogue of 7167 OCs, which is the largest homogeneous OC catalogue up to date.

The astrophysical parameters of OCs have also been determined on a larger scale and with higher precision in the Gaia era. While the ages of small samples of OCs have been measured with asteroseismology \citep{Palakkatharappil23} and gyrochronology \citep{angus2019}, the isochrone fitting method enables the determination of ages for numerous clusters in large amounts, while relying solely on Gaia photometric data. \citet{cantat2020} (hereafter CG20) and HR23 determined the ages using artificial neural networks (ANN) that were trained on synthetic clusters generated from PARSEC isochrones, while other studies applied isochrone fitting via least squares fitting \citep{sim2019} or an MCMC-based method \citep{bossini2019} (hereafter B19).

Isochrone fitting is a multi-parametric process during which the age, metallicity, distance modulus and extinction of a cluster are determined simultaneously. An additional factor is the age-metallicity degeneracy, which introduces an additional uncertainty to the age determination. B19 calculated that a [Fe/H] uncertainty of 0.06 dex leads to an \logage\ uncertainty of 0.01 dex, whereas a [Fe/H] uncertainty of 0.3 dex in turn results in the \logage\ error of 0.05 dex. To limit the age uncertainty they fixed the [Fe/H] value obtained from high-resolution spectra of the clusters, if the spectra were available. However, not all known OCs have this luxury.

\begin{figure}
	\centering
	\includegraphics[width=1\linewidth]{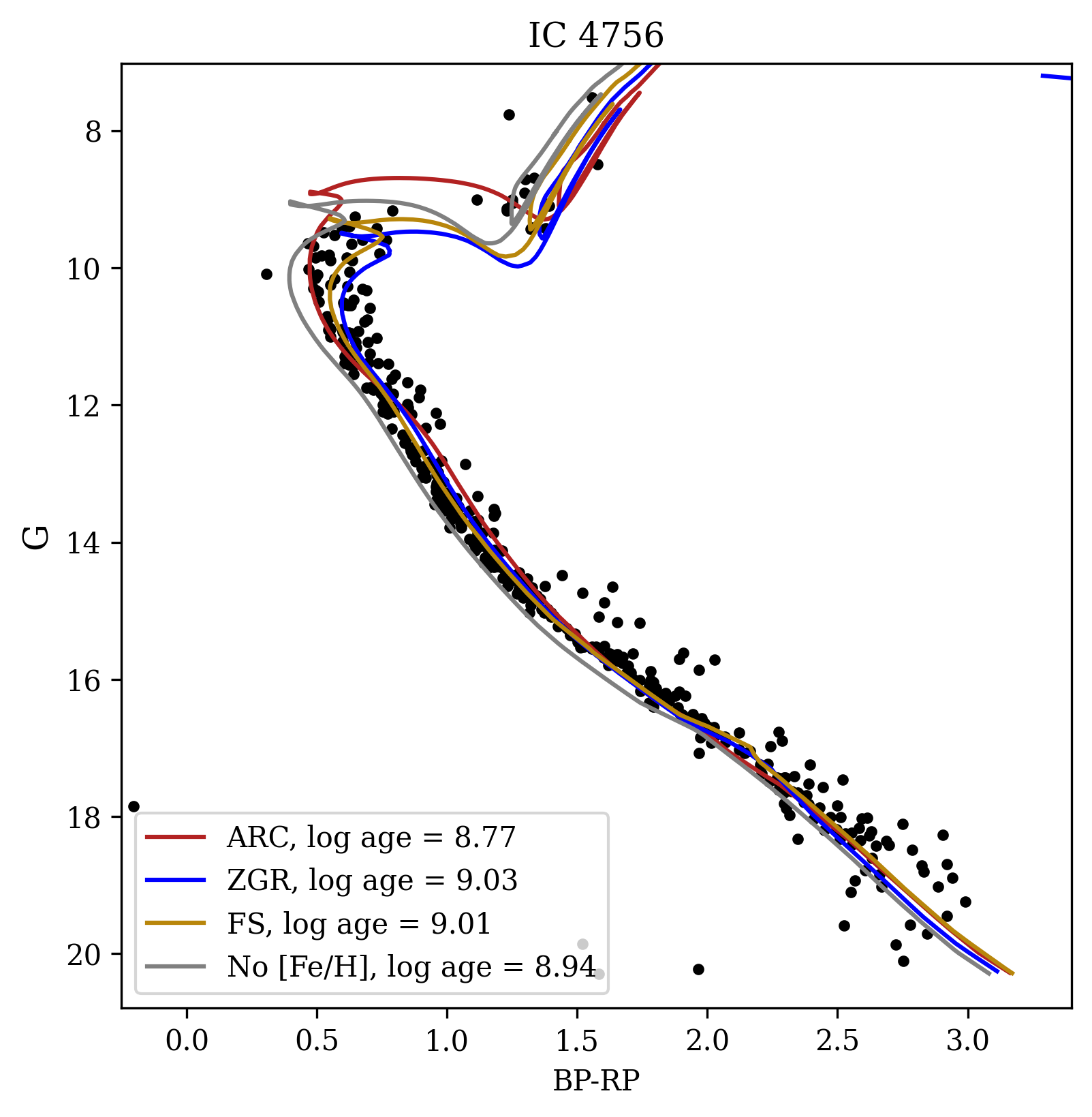}
	\caption{CMD of the open cluster IC 4756 from Gaia DR3 photometry. The isochrones correspond to the derived parameters based on the metallicity prior from the respective Gaia XP-based stellar parameters estimates catalogue (see Section \ref{sect:metallicity} for the description of the catalogues), namely ARC (red), ZGR (blue) and FS (yellow). The grey isochrone corresponds to the fit performed with metallicity as a free parameter.}
	\label{fig:cmd}
\end{figure}
The expansion of Gaia data continued further, as Gaia DR3 \citep{gaiadr3} has introduced the XP spectra for 220 million objects. XP spectra are low-resolution spectra obtained in blue and red Gaia passbands (BP, RP). Despite the low resolution ($R\sim20$--$70$) of these spectra, a growing number of recent studies were able to provide estimates for stellar parameters using machine learning algorithms. \citet{zhang2023} used a forward model to provide estimates for effective temperature (\teff), surface gravity (\logg), and metallicity [Fe/H] for the whole XP dataset, whereas \citet{andrae2023} estimated  \teff, \logg\ and [M/H] for 220 million objects using the XGboost machine learning algorithm. \citet{fs24} were able to determine not only these stellar parameters, but also [C/Fe], [N/Fe] and [$\alpha$/Fe]. \citet{lucey2023} also applied XGBoost to the XP spectra to search for carbon-enhanced stars and provided probabilities for each object being carbon-enhanced. Moreover, \citet{kane2025} identified $\sim$900 candidates of escapees from globular clusters by determining [Al/Fe] and [N/O] from Gaia XP spectra using a trained neural network. Other examples of XP spectra exploitation with machine learning methods include studies by \citet{rvs-cnn, aspgap, yao2024, khalatyan2024, ye2025, hattori2025}, which show the growing interest in extracting chemical information from Gaia data.

In this paper we aim to apply metallicities derived from Gaia XP spectra to the parameter derivation of Milky Way open clusters and test the possibilities of studying OCs with Gaia data alone. Having an estimate of metallicity will help lift the age-metallicity degeneracy, therefore providing more accurate age estimates. We use a selection of clusters previously studied with high-resolution spectroscopy and revisit them with the context of new Gaia data to compare how these age estimates change.

In Section \ref{sect:data} we describe the selection of clusters and the determination of metallicities, and in Section \ref{sect:params} we describe the process of the clusters parameters determination. Section \ref{sect:results} presents the obtained results of dynamical properties for the selected clusters. We discuss the reliability of the input data and the outcome results in Section \ref{sect:discussion}, and we present the overview of our work in Section \ref{sect:conclusions}.
\begin{figure}
	\centering
	\includegraphics[width=1\linewidth]{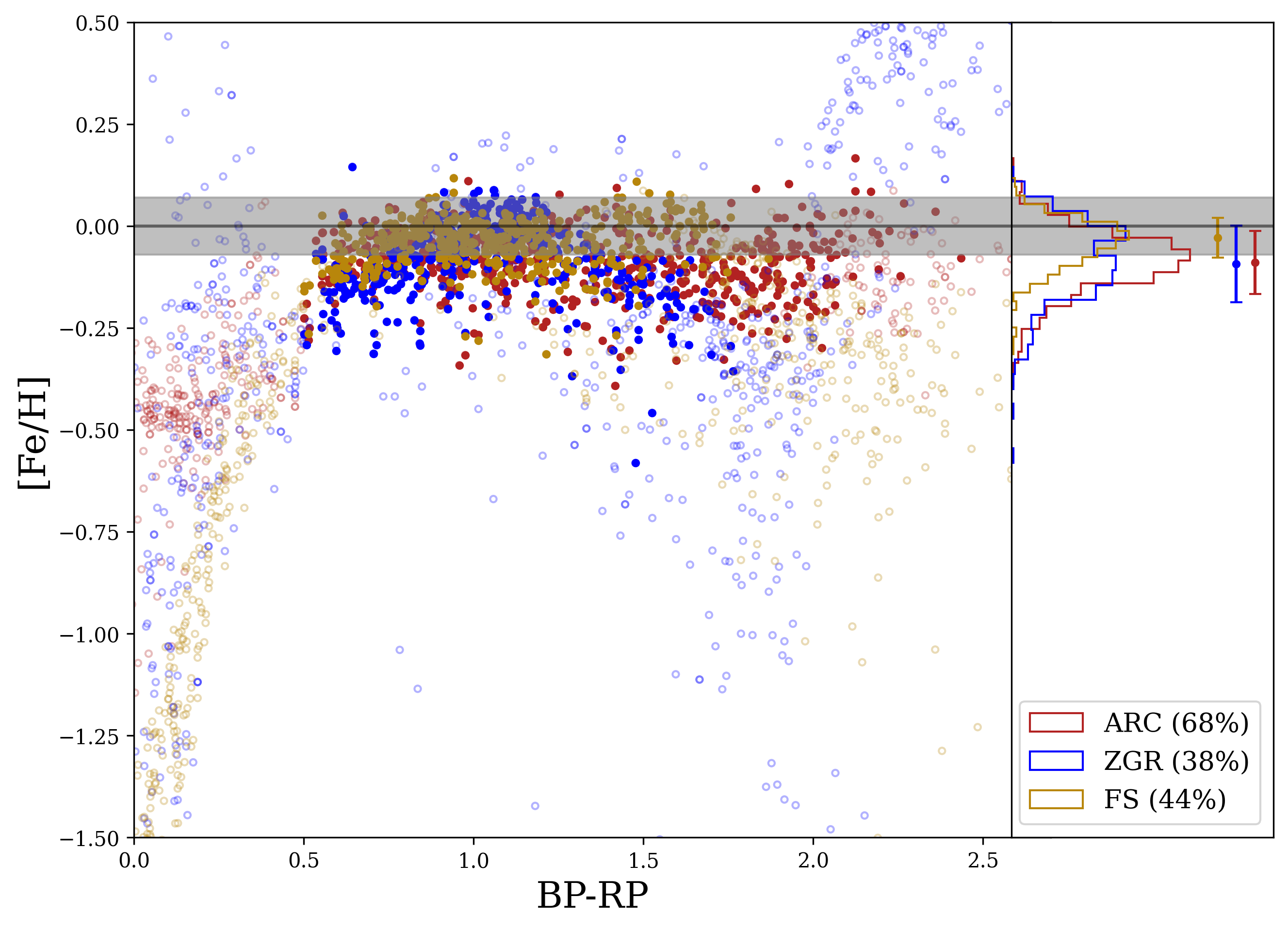}
	\caption{Metallicities determined from the catalogues ARC, ZGR and FS of the open cluster NGC 3532 as a function of Gaia \bprp\ colour. Empty circles show the estimates that were discarded after the respective cuts, whereas filled circles show the values that satisfied the cuts' criteria. The black line and the shaded area show the [Fe/H] and $\sigma$[Fe/H] determined with high-resolution spectroscopy by \cite{Netopil16} and used by B19, whereas the points with error bars show the weighted mean [Fe/H] and its dispersion of the respective catalogue. The percentage indicates the fraction of the stars that were accepted and used for [Fe/H] determination after the respective cuts.}
	\label{fig:bprp_met}
\end{figure}
\section{Data selection} \label{sect:data}
\subsection{Cluster membership}
We started our sample selection from a subset of 37 clusters from B19 with high-resolution spectroscopy (HRS) measurements of metallicity. Their sample was defined to include clusters that have extinction less than $A_V < 2.5 \, \text{mag}$ and are older than 10 Myr ($\log(\text{age}) > 7 \, \text{dex}$) based on the values from \citet{dias2002} and \citet{kharchenko2013}. The latter avoids the potential uncertainty in the age determination that comes with fitting isochrones to young clusters without a clearly defined main-sequence turnoff (MSTO). The [Fe/H] measurements were adopted from \citet{Netopil16} that includes measurements for 88 OCs derived from high-quality, high-resolution (R $\geq$ 25000), and high signal-to-noise ratio (S/N $\geq$ 50) spectra. These spectra were obtained from various literature sources and the results were homogenised. This selection allows us to compare the effect of using low-resolution spectroscopy measurements for age determination. 

For isochrone fitting to the cluster CMDs, we used Gaia DR3 photometry due to its all-sky coverage and exceptional precision at the mmag level for the brighter stars. It is available in three bands: broad $G$ band (330–1050 nm), $G_{BP}$ band (330–680 nm), and $G_{RP}$ band (630–1050 nm). 

To select cluster members, we used the cluster membership probabilities (MP) derived by \citet{hunt2023}. Their catalogue includes stars up to $G = 20 \, \text{mag}$, in contrast to \citet{cantat2018} which is limited to $G = 18 \, \text{mag}$. As a result, the \citet{hunt2023} catalogue contains almost two times as many cluster member stars. Additionally, this catalogue was compiled using not only a clustering algorithm (HDBSCAN) but also a Bayesian convolutional neural network trained on cluster CMDs to verify cluster memberships. In our analysis, we only included cluster members with MP $> 0.5$. An example CMD is shown in Figure \ref{fig:cmd}.

After initial testing, we noticed that in clusters with a heavily populated bottom of the main sequence, the fit was being skewed, thus diminishing the goodness of fit to the stars in the area of the MSTO and the red clump. To accommodate for that, we imposed a magnitude cut on several clusters: G $<$ 15 mag for NGC 2632, G $<$ 16 mag for NGC 6475, and G $<$ 17 mag for NGC 2516 and NGC 3532. 
\begin{figure*}[ht!]
	\centering
	\includegraphics[width=1\linewidth]{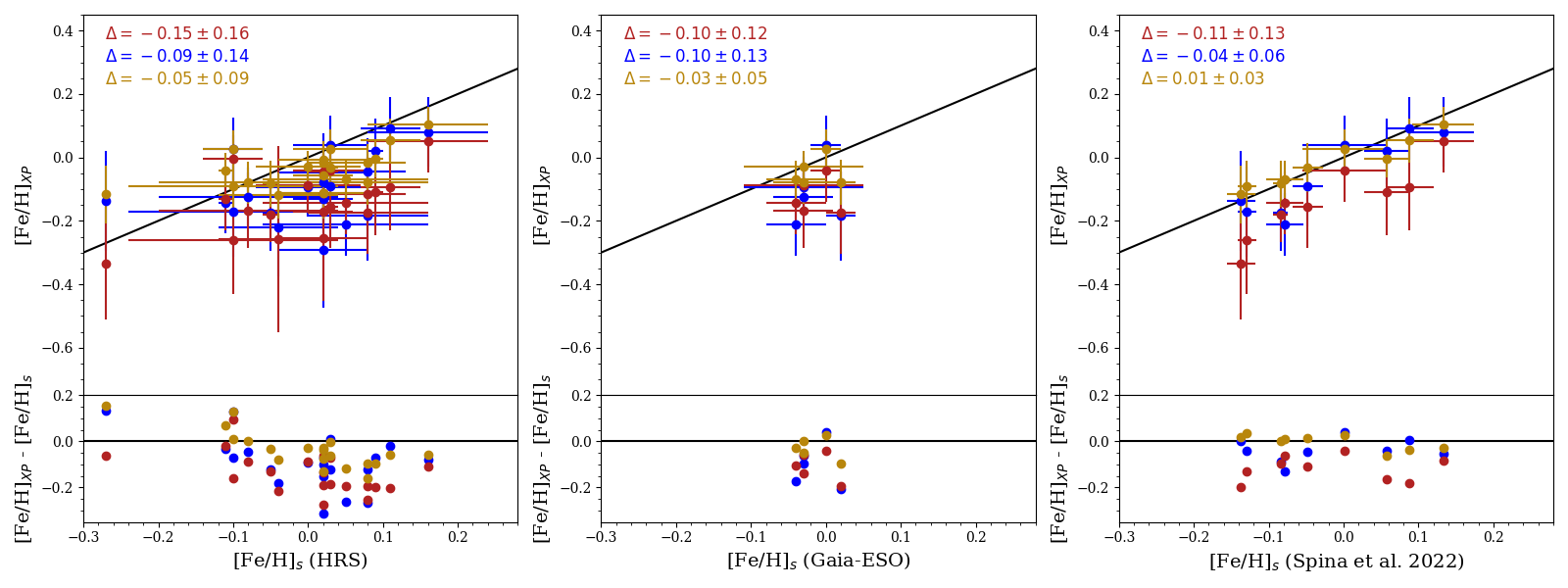}
	\caption{Top panels: Metallicity of individual OCs of the sample derived from three Gaia XP-based catalogues (on the Y axis) presented against that from three high-resolution spectroscopic surveys (on the X axes of the left, middle and right plots). Bottom: Difference between the two values for each cluster. The same colour-coding scheme applies as in Fig. \ref{fig:cmd}. The 1:1 relation is shown by the diagonal black line in the top plot and, respectively, by the horizontal line indicating zero difference in the bottom plot. The number in the top corner indicates the median difference and the root mean square difference (RMSD) of the respective catalogue.}
	\label{fig:met_compare} 
\end{figure*}
\subsection{Metallicity} \label{sect:metallicity}
\begin{figure}[ht!]
	\centering
	\includegraphics[width=1\linewidth]{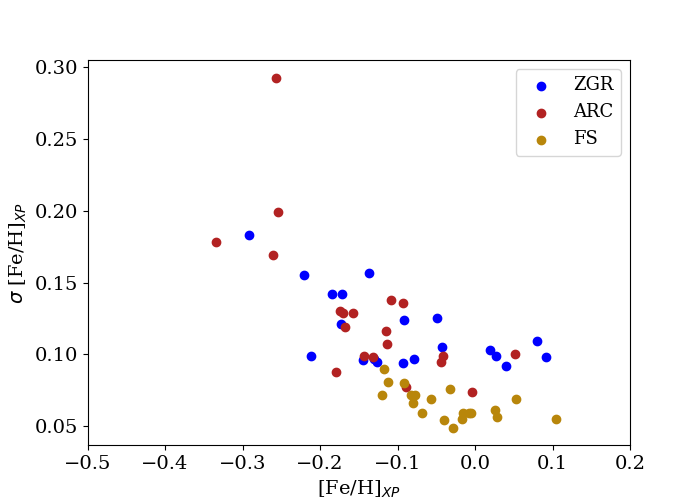}
	\caption{Iron abundance dispersion $\sigma \text{[Fe/H]}$ against metallicity [Fe/H] for individual open clusters. The same colour-coding scheme applies as in Fig. \ref{fig:cmd}.}.
	\label{fig:met_errors}
\end{figure}
To calculate the metallicities of the open clusters, we used measurements based on Gaia XP spectra from several recently published catalogues. Using Gaia XP spectra instead of the higher resolution Gaia RVS spectra allowed us to increase the number of measurements per cluster, since the number of stars with metallicities derived from RVS spectra is 5.6 million in contrast to 220 million of XP spectra.  Due to the very low resolution of the latter, it becomes problematic to determine the stellar parameters using standard methods such as spectral fitting or equivalent width measurements. Therefore, studies utilise machine learning algorithms that are trained on spectroscopic libraries of high-resolution surveys in order to extract the stellar parameters information from the spectra.

We used the results from the following catalogues:

\textit{ZGR} \citep*{zhang2023} contains approximately 220 million stars. The empirical forward model was trained on six million LAMOST spectra (R $\sim$ 1800) and augmented with 2MASS and WISE photometry. Deliverables of the catalogue include temperature (T$_{\text{eff}}$), surface gravity ($\log g$), metallicity ([Fe/H]), parallax ($\varpi$), and extinction (E). They report a typical error of [Fe/H] measurement to be 0.15 dex. We applied the recommended quality cut that denotes that all three parameters (T$_{\text{eff}}$, $\log g$, and [Fe/H]) are reliable: \texttt{quality\_flags < 8}. An example of this cut is seen in Figure \ref{fig:bprp_met}.

\textit{ARC} \citep*{andrae2023} provides metallicity [M/H], as well as T$_{\text{eff}}$ and $\log g$, for 175 million stellar objects. This is an improvement upon the previous work of \citet{rix2022} that did this work for two million bright giant stars. ARC utilised the XGBoost algorithm that was trained on 600,000 APOGEE DR17 \citep{apogee} spectra (S/N $>$ 50) and supplemented with CatWISE photometry \citep{catwise}. The model covers the temperature range from 3107 K to 6867 K, and the metallicity range from -4.37 to +0.5 dex. The training sample includes main-sequence dwarfs and red giant stars, but there is limited coverage for [M/H] $<$ -2 dex and \teff $<$ 4000 K, as well as very few training examples for metal-poor dwarfs with \logg $>$ 3.5 dex and [M/H] $<$ -1 dex. 
They achieved a precision of $\sigma(\text{[M/H]}) = 0.1$ dex, making it outstanding quality data for its size. Following the prescriptions of the study, we applied several selection cuts:
\begin{itemize}
    \item $0.5 < G_{BP}$ - $G_{RP} < 2.5$,
    \item $G < 17$ mag,
    \item filtering out the OBA stars based on the Gaia Golden OBA sample \citep{gaiadr3}.
\end{itemize}

\textit{FS} \citep*{fs24} present the predictions of a machine learning model for stellar atmospheric parameters of using Gaia BP/RP spectra and broadband photometric colours from Gaia, 2MASS, and WISE. The model predicts six stellar parameters: \teff, \logg, iron [Fe/H], carbon [C/Fe], nitrogen [N/Fe], and alpha [$\alpha$/M], along with uncertainties for these parameters.  It was trained on a dataset of APOGEE  spectra, which includes over 650,000 stars. The model's predictions were validated against APOGEE and LAMOST spectroscopic data, and Gaia GSP-Phot parameters. The quality of the deliverables, particularly for metallicity ([Fe/H]), is reported with a median formal uncertainty of 0.068 dex, indicating high precision in the metallicity predictions. While there are no explicitly recommended cuts for this catalogue, we applied the following limitations:
\begin{itemize}
    \item $G_{BP}$ - $G_{RP} > 0.5$,
    \item $\sigma$ [Fe/H] $< 0.1$.
\end{itemize}

The first cut was dictated by a sharp decline in [Fe/H] for the hottest stars of the MS observed in almost every cluster in the sample (for example, see Figure \ref{fig:bprp_met}) and is caused by the model being undertrained with OB stars. The second cut was based on the reported median [Fe/H] uncertainty of the catalogue. 

Other catalogues of metallicities based on Gaia XP spectra were considered:

\textit{GSP-Phot} \citep{gsp-phot} is a module in the Gaia processing pipeline that aims to determine stellar parameters using only Gaia photometry, parallaxes, and BP/RP spectra. However, the [M/H] values exhibit strong systematics and are not recommended for use by the Gaia collaboration. Moreover, \citet{bianchini2024} showed the influence of these systematics in the globular cluster NGC 5904, which was previously reported to host populations of two metallicities based on the GSP-Phot data. Comparisons with high-quality ARC metallicity values revealed that the cluster is, in fact, monometallic.

\textit{Pristine} \citep{pristine-gaia} presented synthetic CaHK-based metallicities for 220 million stars by training a neural network on CaHK metallicities derived from the Pristine survey. We do not use this  otherwise extensive catalogue since it was calibrated for metal-poor stars ([Fe/H] $<$ -0.3 dex), which is lower than the metallicity values of the clusters in our sample.

For the analysis of the HRS sample, we only considered open clusters with more than 100 measurements in both the \citet{zhang2023}, \citet{andrae2023} and \citet{fs24} catalogues in order to ensure the statistical significance of the derived metallicities. After applying these cuts, our final sample consisted of 20 open clusters. 
\section{Cluster parameters} \label{sect:params}
We used the ASteCA (Automated Stellar Cluster Analysis) package \citep{asteca} to analyse the cluster properties. It is based on the ‘synthetic CMD method’, in which a synthetic cluster is generated based on a provided isochrone and cluster parameters, and then compared to the observational data. The best model in this process is determined with Bayesian statistics and is described by four parameters: [Fe/H], age, distance modulus and extinction $A_{V}$. We used the 0.5.8 version of the code, which supports the use of approximate Bayesian computation (ABC) via the implementation of pyABC \citep{pyabc}. 

As input for the code, we used the $G$ and \bprp\ magnitudes, their corresponding errors, and a set of PARSEC 1.2S\footnote{\url{http://stev.oapd.inaf.it/cgi-bin/cmd}} isochrones in the Gaia EDR3 photometric system. For the isochrones, we used a grid of 7.2 dex $<$ log(age) $<$ 10.1 dex with a step of 0.01 dex, and -2.1 $<$ [Fe/H] $<$ 0.5 with a step of 0.05 dex. 

Although ASteCA includes a decontamination algorithm which then assigns a membership probability to each source, we used the MPs provided with \citet{hunt2023} catalogue. To populate the isochrone of a synthetic cluster we used the initial mass function (IMF) derived by \citet{kroupa2002}. In contrast to the BASE-9 code \citep{base9}, which was used by B19, ASteCA allows one to include binary systems in the synthetic cluster. To accommodate for the binary distribution we used the standard values of $\alpha = 0.5$ and $\beta = 0.15$, where the probability of a star being in a binary system is determined by the formula:
\begin{equation}
P = \alpha + \beta \text{log}_{10}(M),
\end{equation}
and where $M$ is the mass of the system expressed in $M_{\odot}$.

We provided the priors for the parameters since it is a necessary component of the Bayesian method. The prior of the distance modulus was provided as a Gaussian function, the standard deviation $\omega$ of which was 0.3 mag, and the mean was specified by utilising the median parallaxes of the cluster available with Gaia via the following equation:
\begin{equation}
	(m-M) = -5\log(\widetilde{\varpi})-5,
	\label{eq:parallax}
\end{equation}
where $(m-M)_0$ is the distance modulus and $\widetilde{\varpi}$ is the median value of the parallax of the stars included as cluster members. 
The prior of the extinction $A_{V}$ was flat and limited to a maximum of 1.4 mag, which is stricter than the 2.5 mag limit used by B19. This choice was made to narrow the parameter space while still safely covering the full range of $A_{V}$ values observed in our selected subsample, based on B19 measurements.
 The prior for age was not limited and the age could vary across the grid values. Lastly, for the prior of metallicity we used one of the following: \\
1) the fixed [Fe/H] value used by B19, \\
2) the derived [Fe/H] values of clusters and their respective $\sigma_{[Fe/H]}$ from the ZGR, ARC, or FS survey, \\
3) no prior provided and the metallicity was a free parameter.

We performed each run using the pyABC algorithm with a population size of 100 particles per iteration and a maximum of 20 iterations allowed. We established the acceptance threshold at $\epsilon = 0.01$, which determines the maximum allowed distance between the simulated and observed datasets for a sample to be accepted. \\
\begin{figure}
	\centering
	\includegraphics[width=0.7\linewidth]{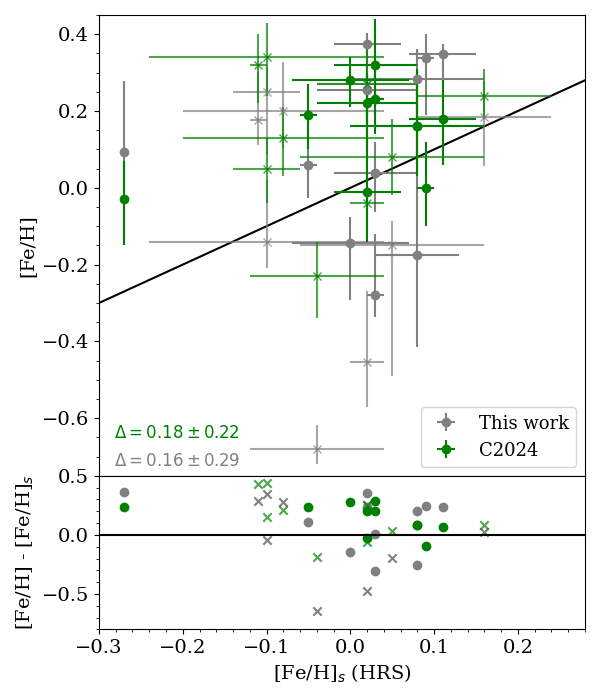}
	\caption{Metallicity of individual OCs of the HRS sample determined as a free parameter in the procedure. The grey colour denotes this work; the green colour shows the predictions of the ANN of \citet{cavallo24}. Figure layout is the same as in Fig. \ref{fig:met_errors}: X axis shows the [Fe/H] from high-resolution spectroscopy, the bottom plot shows the difference between the values, and the black line indicates 1:1 relation on the top plot and difference of zero in the bottom plot. Clusters with three or fewer red giant branch (RGB) stars are marked with an X symbol. The number in the bottom left corner indicates the median difference and RMSD.}.
	\label{fig:met_algorithms}
\end{figure}
In order to correctly accommodate for extinction in Gaia photometric bands during the isochrone fitting process, we utilised the conversion of coefficients a$_{..m}$ from Gaia (E)DR3 auxiliary data\footnote{\url{https://www.cosmos.esa.int/web/gaia/edr3-extinction-law}} available in ASteCA:
\begin{align}\label{eq:av-conversion}
	\frac{A_m}{A_V} &= a_{1m} + a_{2m} X + a_{3m} X^2 + a_{4m} X^3 + a_{5m} A_V + a_{6m} A_V^2 \notag \\
	&\quad + a_{7m} A_V^3 + a_{8m} A_V X + a_{9m} A_V X^2 + a_{10m} X A_V^2,
\end{align}
where $m$ is $G$, $G_{BP}$ or $G_{RP}$, and X is $G_{BP} - G_{RP}$.

\begin{figure*}[ht!]
	\centering
	\includegraphics[width=1\linewidth]{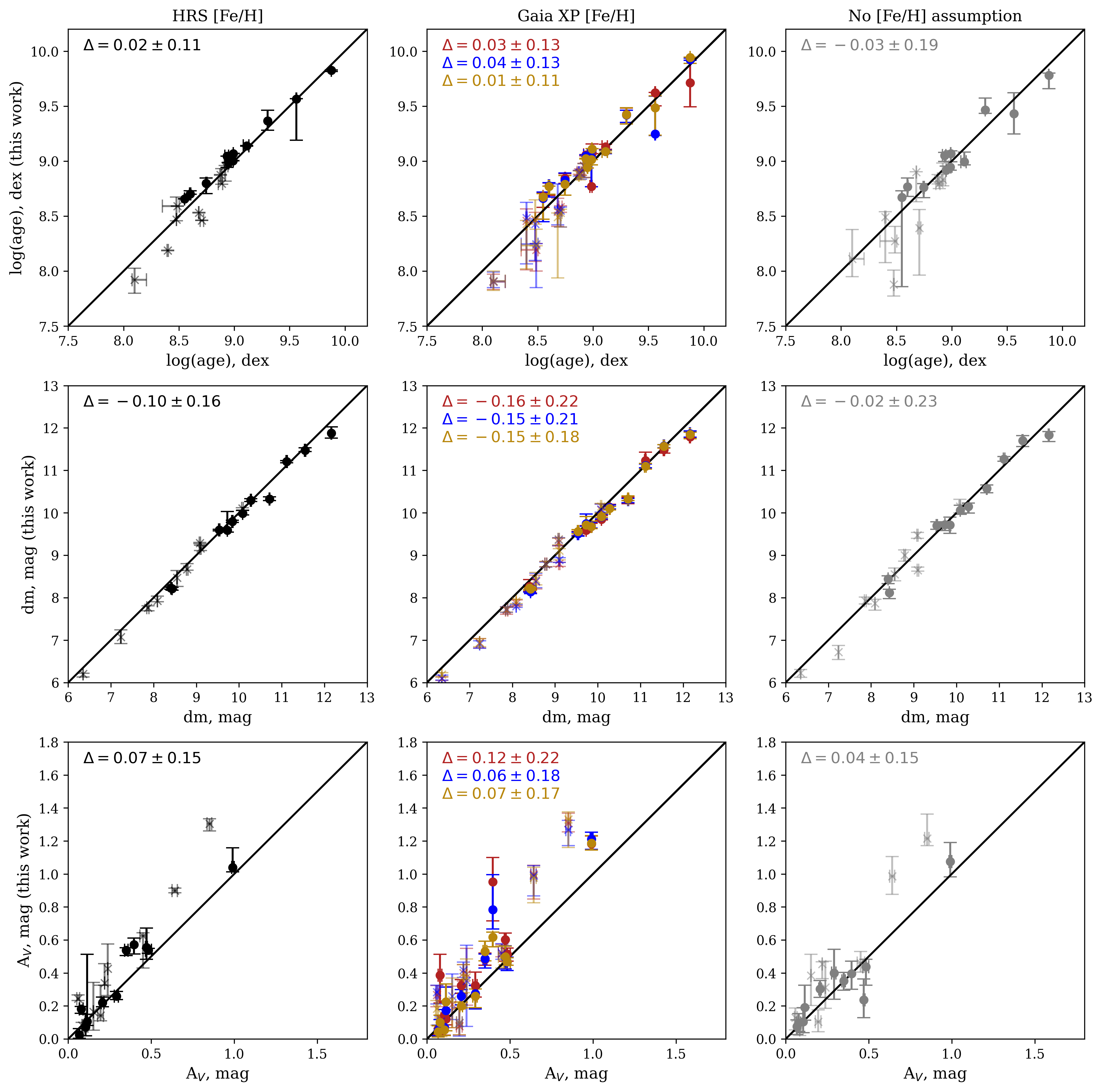}
	\caption{Parameter estimates for age, distance modulus and extinction (rows, Y axis) of the OCs determined with different [Fe/H] prior information (columns) compared to those from B19 on the X axis. The red, blue, and yellow colours correspond to runs with metallicity prior based on ARC, ZGR and FS catalogues, respectively, whereas black and grey symbols, respectively, depict the runs with the fixed HRS [Fe/H] prior and no prior at all. The legend in the upper corner of the figures shows the median absolute difference and the RMSD of the respective parameter. The OCs with three or fewer RGB stars are marked with an X symbol.}
	\label{fig:age-compare}
\end{figure*}
\section{Results} \label{sect:results}
\subsection{Comparison of [Fe/H] with spectroscopic surveys} \label{sect:feh}
First, we investigated the distribution of [Fe/H] measurements from each catalogue in each of the clusters of the sample. The median rate of stars in each cluster that have measurements in the ARC, ZGR, and FS catalogues is 72\%, 84\%, and 75\%, respectively. After the cuts applied to these catalogues, the final number of stars for which the [Fe/H] measurements are accepted drops to a median of 50\%, 49\%, and 38\%, respectively. The difference of $\sim10\%$ between the FS catalogue and the other two already demonstrates the resulting strictness of the cuts applied to the values of FS.

For every cluster in the sample, we calculated the weighted mean metallicity value from the ZGR and FS catalogues, using the provided errors. For the ARC catalogue, as it does not contain information about errors, we only calculated the mean metallicity. While these surveys provide estimates of either [M/H] (ARC) or [Fe/H] (ZGR, FS), in the further analysis we refer to both of them as [Fe/H], since the standard deviation of these values exceeds the typical alpha-enhancement.
\begin{figure}
	\centering
	\includegraphics[width=1\linewidth]{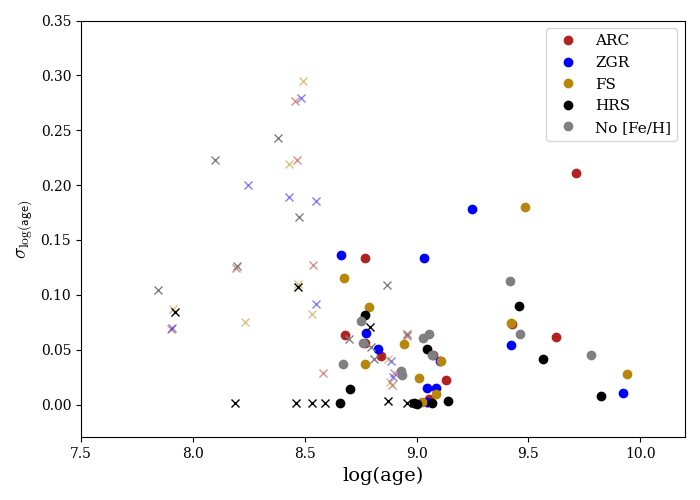}
	\caption{Error of the fitted age parameter, calculated from the 84th and 16th percentiles of the posterior distribution, \siglogage\ against the \logage\ (50th percentile) for individual open clusters of the sample. The same colour-coding scheme applies as in Fig. \ref{fig:age-compare}.}
	\label{fig:age-err}
\end{figure}

To assess the reliability and potential biases of these measurements, we compared them with those obtained from high-resolution spectroscopic surveys and used in the HRS sample of B19 (see left panel of Figure \ref{fig:met_compare}). According to these estimates, the [Fe/H] of the cluster sample derived from Gaia XP-based catalogues is systematically lower than HRS values. This difference varies between the three catalogues that were used. The median difference between HRS and ARC [Fe/H] estimates equals to $-0.15 \pm 0.16$ dex, comparable to $-0.09 \pm 0.14$ dex difference between HRS and ZGR [Fe/H] values. The FS catalogue, in turn, shows a far better agreement with the HRS values. The median difference between HRS and FS is $-0.05 \pm 0.09$ dex.

While the HRS sample of B19 is based on high-resolution spectroscopic measurements from \citet{Netopil16}, the number of measured stars per cluster is quite limited. Namely, 15 out of 20 clusters in our sample have <5 measurements in \citet{Netopil16}. This is in stark contrast with the average number of 350 stars from the ARC and ZGR catalogues and 250 stars from the FS catalogue, after applying the cuts. To ensure the robustness and consistency in our results, we compared them to the more recent and extensive studies of the metallicity of open clusters with high-resolution spectroscopy: Gaia-ESO and \citet{spina2022}. 

The Gaia-ESO survey (GES, \citet{gaia-eso-oc-mapping, gaia-eso-randich}) intended to complement the astrometry of Gaia with high-precision radial velocities and abundances using both GIRAFFE ($R \sim 20 000$) and UVES ($R \sim 47 000$) spectrographs. A part of the survey was specifically dedicated to open clusters and comprised 62 clusters accompanied by ESO archival spectra for another 18 clusters, in total covering an extensive range of ages (0.001--10 Gyr). The main sample has been observed with both spectrographs, whereas the clusters in the ESO archive only have UVES spectroscopy available. We found only five open clusters in our sample that overlap with the GES OC survey: NGC 2516, NGC 3532, NGC 6633, and two archival clusters, NGC 5822 and NGC 2682. While the number of measurements for non-archival data is greater than 700 per cluster, the archival clusters have 4 and 131 measurements, respectively. In total, the average error $\sigma(\text{[Fe/H]})$ of this sample is 0.04 dex, which is indicative of the high quality of the spectra and benefits from the large number of measurements per cluster. However, this level of precision is available for only a small number of clusters, limiting its broader applicability.

Despite covering a narrower range of [Fe/H], the results shown in middle panel of Figure \ref{fig:met_compare} are similar to the previous comparison with B19 both in terms of accuracy and precision: [Fe/H] derived from the FS catalogue persistently shows excellent agreement and an RMSD of 0.05 dex. In contrast, the other two catalogues differ from Gaia-ESO measurements by approximately 0.1 dex.

Given the limited overlap of our cluster sample with GES, we also compared the derived metallicities with the catalogue of the metallicity of 251 OCs by \citet{spina2022} (see right panel of Figure \ref{fig:met_compare}). They traced the Galactic metallicity gradient by using homogenised [Fe/H] from high-resolution spectroscopic surveys: APOGEE, GES, GALAH, OCCASO, and SPA. We find that 9 out of 20 clusters from our sample have metallicities in \citet{spina2022}. From this selection, the results are consistent with the previous comparisons, as ARC shows a noticeable systematic difference of approximately $-0.1$ dex, and FS shows a small RMSD of less than 0.05 dex.

The standard error $\sigma(\text{[Fe/H]})$ exhibits a clear [Fe/H] dependence (Fig.\ref{fig:met_errors}) and is more pronounced in ZGR and ARC catalogues (see Section \ref{sect:discussion}). The median $\sigma(\text{[Fe/H]})$ derived from these catalogues for an individual cluster are 0.12 dex and 0.13 dex, respectively, which are comparable to the reported precision of 0.15 dex and 0.10 dex for the whole catalogues. FS shows a smaller median $\sigma(\text{[Fe/H]})$ of 0.07 dex, which once more reflects the quality of the data combined with the imposed maximum of 0.1 dex for uncertainties for [Fe/H] measurements of individual cluster members.

\subsection{[Fe/H] as a free parameter}\label{section:feh-free}
One of the three regimes in which we determined the clusters' parameters is by not providing a [Fe/H] prior at all, leaving it as a free parameter, and allowing it to range across the whole grid. This test was done in order to evaluate how, if at all, necessary the [Fe/H] prior is and how its accuracy in turn affects other parameters. While we do discuss these parameters in the next section, this regime does leave the [Fe/H] estimates found from the best fit as corollary.

As is clearly seen in Figure \ref{fig:met_algorithms}, the [Fe/H] obtained as a parameter in the CMD fitting shows considerably less accurate results than when it is derived from Gaia XP-based catalogues. When compared to the [Fe/H] from the HRS sample (shown on the x axis), these values have a significantly larger RMSD of 0.29 dex, which is nearly double the RMSD achieved using XP-based [Fe/H], which is detailed in Section \ref{sect:feh}. This demonstrates that leaving metallicity unconstrained in the fit leads to degraded performance. 

We also compared these results to a neural network-based (NN) study, where [Fe/H] was also determined as a free parameter. Recent study of \citet{cavallo24} has shown how to determine the parameters of 5400 open clusters, along with determining the metallicity. They used an approach in which the information about a cluster is extracted in the form of features from the photometry using the QuadTree algorithm, and these features are in turn passed to the ANN trained on a grid of synthetic clusters. We refer to the original paper for a more detailed explanation of the technique.

While the method of \citet{cavallo24} allows for the retrieval of some [Fe/H] information from the photometry of open clusters without using the spectra, it has shown an RMSD of 0.35 dex when verified against the study of \citet{dias2021} and was reported to show ‘mild correlation’ with the values of \citet{spina2022}. For our sample, we extracted the [Fe/H] values from \citet{cavallo24} and compared them to the HRS data: the resulting RMSD is 0.22 dex, which is similar to the free-fit approach but also shows a large systematic offset of $\sim$0.2 dex, with [Fe/H] of some clusters overestimated by as much as 0.5 dex. This result highlights that XP-derived metallicities, despite the low resolution of the spectra, provide far better constraints on [Fe/H] than fitting it as a free parameter or relying solely on photometric machine-learning estimates. 

\subsection{Comparison of cluster parameters with B19}
The values of \logage, distance modulus (dm) and extinction \(A_V\) that were determined for the clusters in the sample are the median values of the respective posterior distribution, whereas the reported uncertainties are the 16th and 84th quantiles. We show these estimates as rows in Figure \ref{fig:age-compare}, where the columns show different [Fe/H] prior assumptions. The median absolute differences and RMSD of these parameters with B19 are presented in Table \ref{tab:age-compare}.

\subsubsection{Age}
Since our sample is based on the HRS sample of B19, we first analysed the cluster parameters that were derived with the [Fe/H] being fixed at the same values as in their analysis (top left panel of Figure \ref{fig:age-compare}). Before analysing the effect of various metallicity assumptions on the determination of ages, we evaluated possible discrepancies that arise due to differences between the methodologies of our work and their study. 

The ages demonstrate a reasonable agreement with the ages of B19, with a median difference of 0.02 dex and an RMSD of 0.11 dex. The clusters that show the biggest difference in the ages are those without a significant red giant population, which in turn tend to be on the younger side of the covered age range (\(\log \text{age} < 8.5~\text{dex}\)). This result is similar in accuracy to how the B19 HRS sample compares to the well-established catalogues of MWSC and DAML, with differences of \(\Delta \log \text{age} = 0.04 \pm 0.10~\text{dex}\) and \(0.05 \pm 0.11~\text{dex}\), respectively (see Table 4 in their work).

While in this run the [Fe/H] prior is the same as in their work, as well as the PARSEC isochrones and the [Fe/H] and age steps of the isochrone grid, the discrepancies between the results can be attributed to the algorithms used for isochrone fitting. We use pyABC as implemented by ASteCA, whereas B19 fit the parameters with MCMC via BASE-9. Moreover, while we allow \(A_V\) to vary across the whole range of values, they use prior \(A_V\) information from DAML and MWSC catalogues and limit the prior within \(\sigma_{A_V} = \frac{1}{3} A_V\).

We then considered how a [Fe/H] based on Gaia XP, instead of high-resolution spectroscopy, alters the age determination of the clusters (top middle panel of Figure~\ref{fig:age-compare}). The results show reasonable agreement and significant dispersion, with the median differences between the ages not exceeding 0.05 dex. The ages derived using the [Fe/H] from the FS catalogue show better agreement with the HRS sample, with an RMSD of 0.11 dex, than those derived with ARC or ZGR values (0.13 dex). The accuracy of the ages using the FS catalogue is equal to the ages derived with metallicities from high-resolution spectroscopy, demonstrating the potential of Gaia XP-based metallicities.

\begin{table*}[]
	\centering
	\caption{Median absolute difference and the RMSD of the parameter estimates determined with different metallicity priors with respect to B19 values.}
	\label{tab:age-compare}
	\begin{tabular}{lll|lll|l}
		\hline
		\hline
		&         & HRS [Fe/H]             & \multicolumn{3}{|c|}{Gaia XP [Fe/H]}  & No [Fe/H] assumption  \\
		&         &                 & ARC             & ZGR             & FS              &                   \\
		\hline
		\multirow{3}{*}{$\Delta$ \logage} 
		& All     & \( 0.02 \pm 0.11\) & \(0.03 \pm 0.13\) & \(0.04 \pm 0.13\) & \(0.01 \pm 0.11\) & \(-0.03 \pm 0.19\) \\
		& RGB     & \( 0.05 \pm 0.07\) & \(0.10 \pm 0.13\) & \(0.10 \pm 0.14\) & \(0.07 \pm 0.09\) & \(0.02 \pm 0.11\) \\
		& No RGB  & \(-0.09 \pm 0.14\) & \(-0.01 \pm 0.14\) & \(-0.05 \pm 0.13\) & \(0.00 \pm 0.14\) & \(-0.09 \pm 0.25\) \\
		\hline
		\multirow{3}{*}{$\Delta$ \(dm\)} 
		& All     & \(-0.10 \pm 0.16\) & \(-0.16 \pm 0.22\) & \(-0.15 \pm 0.21\) & \(-0.15 \pm 0.18\) & \(-0.02 \pm 0.23\) \\
		& RGB     & \(-0.10 \pm 0.18\) & \(-0.14 \pm 0.22\) & \(-0.15 \pm 0.20\) & \(-0.17 \pm 0.19\) & \(-0.03 \pm 0.17\) \\
		& No RGB  & \(-0.10 \pm 0.13\) & \(-0.17 \pm 0.21\) & \(-0.16 \pm 0.21\) & \(-0.12 \pm 0.17\) & \(0.00 \pm 0.29\) \\
		\hline
		\multirow{3}{*}{$\Delta$ \(A_V\)} 
		& All     & \(0.07 \pm 0.15\) & \(0.12 \pm 0.22\) & \(0.06 \pm 0.18\) & \(0.07 \pm 0.17\) & \(0.04 \pm 0.15\) \\
		& RGB     & \(0.05 \pm 0.09\) & \(0.12 \pm 0.21\) & \(0.02 \pm 0.14\) & \(0.02 \pm 0.11\) & \(0.01 \pm 0.09\) \\
		& No RGB & \(0.17 \pm 0.21\) & \(0.16 \pm 0.23\) & \(0.11 \pm 0.22\) & \(0.12 \pm 0.22\) & \(0.07 \pm 0.21\) \\
		\hline
		\multirow{3}{*}{$\Delta$ \(\text{[Fe/H]}\)} 
		& All     & --- & \(-0.15 \pm 0.16\) & \(-0.09 \pm 0.14\) & \(-0.05 \pm 0.09\) & \(0.16 \pm 0.29\) \\
		& RGB     & --- & \(-0.19 \pm 0.18\) & \(-0.12 \pm 0.16\) & \(-0.06 \pm 0.09\) & \(0.20 \pm 0.25\) \\
		& No RGB  & --- & \(-0.11 \pm 0.13\) & \(-0.07 \pm 0.13\) & \(-0.03 \pm 0.07\) & \(0.03 \pm 0.34\) \\
	\end{tabular}
	\tablefoot{The results for all clusters are also shown in Fig. \ref{fig:age-compare}, and the results of the Gaia XP metallicities are shown in Fig. \ref{fig:met_compare}. They are also separated into two groups, based on the presence of RGB stars (‘RGB’: $>$3 stars, ‘No RGB’: $\leqslant$3 stars).}
\end{table*}

Lastly, we inspected the ages derived from the isochrone fits with no [Fe/H] prior, shown in the top right panel of  Figure~\ref{fig:age-compare}. Since the [Fe/H] derived from these fits has a substantial RMSD (see Section~\ref{section:feh-free}), the age estimates also show a considerably larger RMSD of \(\sim0.2~\text{dex}\) in contrast to the fits with [Fe/H] prior information. The main contributors to this dispersion are young clusters with no defined post-main sequence (MS) stars, as expected, whereas older clusters show a small difference in age. This is seen in Table~\ref{tab:age-compare}, where the ‘No RGB’ subgroup exhibits the largest RMSD of 0.25 dex  in \logage\ when no metallicity prior is used while clusters with RGB stars remain largely unaffected by using different metallicity priors. These findings suggest that the inclusion of a metallicity prior is particularly beneficial for clusters lacking evolved stars, as it significantly enhances the accuracy of their age estimates.

The comparisons outlined above establish the accuracy of the ages that we derived. In order to assess their precision, we calculated the median \siglogage\ from the individual age estimates (see Fig.~\ref{fig:age-err}). The median \siglogage\ of the ARC, ZGR, and FS catalogues is \(0.062~\text{dex}\), \(0.065~\text{dex}\), and \(0.069~\text{dex}\), respectively. If we only take into consideration the clusters with >3 post-MS stars, the respective average \siglogage\ decreases to \(0.056~\text{dex}\), \(0.050~\text{dex}\), and \(0.040~\text{dex}\). This is comparable to the median \siglogage\ for estimates obtained without a [Fe/H] prior (\(0.056~\text{dex}\)). The precision of the ages derived with the HRS [Fe/H] is \(0.008~\text{dex}\), which is an order of magnitude lower and is dictated by the fixed value of [Fe/H], which in turn reduces the degeneracy between the fit parameters. 

\subsubsection{dm and $A_V$} \label{sect:dm-Av}
We also compared the other two parameters of the fitting procedure: distance modulus and extinction $A_V$ (middle and bottom rows of Figure~\ref{fig:age-compare} respectively). When derived from fits with different [Fe/H] assumptions, distance modulus demonstrates the same trend of dispersion in comparison to the values of B19, as the fits with Gaia XP-based [Fe/H] show bigger dispersion than with fixed HRS [Fe/H] (0.16 mag) but smaller than with no [Fe/H] prior at all (0.23 mag). Fits that used the FS catalogue have a smaller RMSD between the three catalogues, in contrast to the fits that used [Fe/H] of the ARC catalogue. Similar to \logage\ estimates, the dm estimates for clusters without RGB stars (see Table \ref{tab:age-compare}) show the largest improvement from including Gaia XP-based priors, as RMSD drops from 0.29 mag without the prior to 0.17 for the FS catalogue. Clusters with RGB stars, however, show minimal improvement of the dm estimate.

As for the extinction $A_V$, it is consistently overestimated in all three types of fits for two clusters with $A_V > 0.6$ mag in B19 and an unpopulated RGB. All three types of fits show a positive offset, with that of the ARC catalogue being the largest. Notably, the RMSD of the $A_V$ of clusters with HRS [Fe/H] and no [Fe/H] prior is the same. The presence of RGB stars has a clear influence on the accuracy of the extinction estimate, as clusters with RGB stars have a significantly lower RMSD (0.09 mag) compared to those without (0.21 mag). However, Gaia XP-based priors do not appear to improve the precision of $A_V$ in either group, in contrast to \logage\ and dm parameters. For clusters with RGB stars, the use of Gaia XP [Fe/H] actually increases the dispersion in extinction estimates. We examine this issue in the next section.

\begin{figure*}[ht]
	\centering
	\begin{subfigure}[b]{0.48\textwidth}
		\centering
		\includegraphics[width=\textwidth]{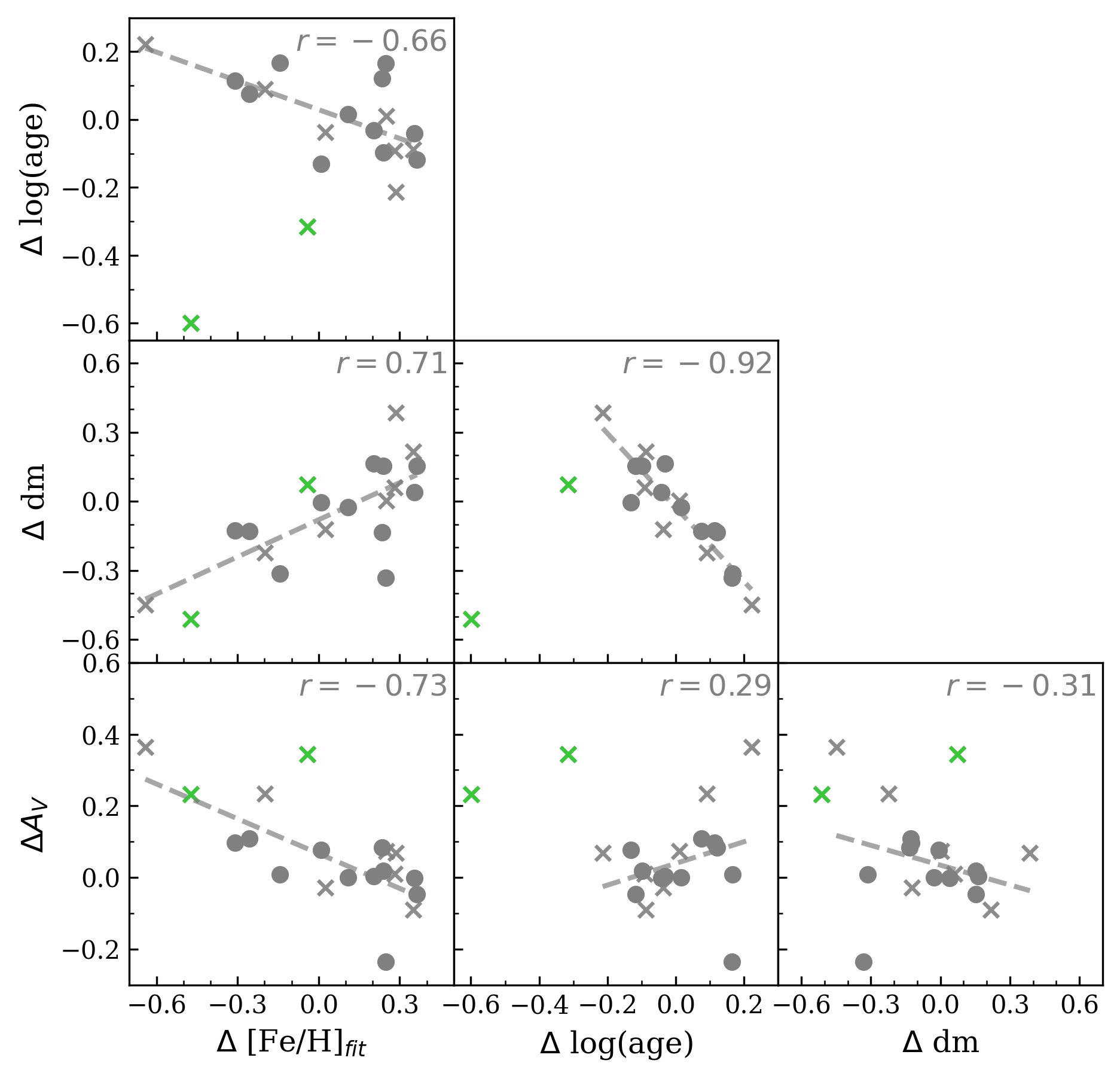}
		\caption{}
		\label{fig:corner_nofix}
	\end{subfigure}
	\hfill
	\begin{subfigure}[b]{0.48\textwidth}
		\centering
		\includegraphics[width=\textwidth]{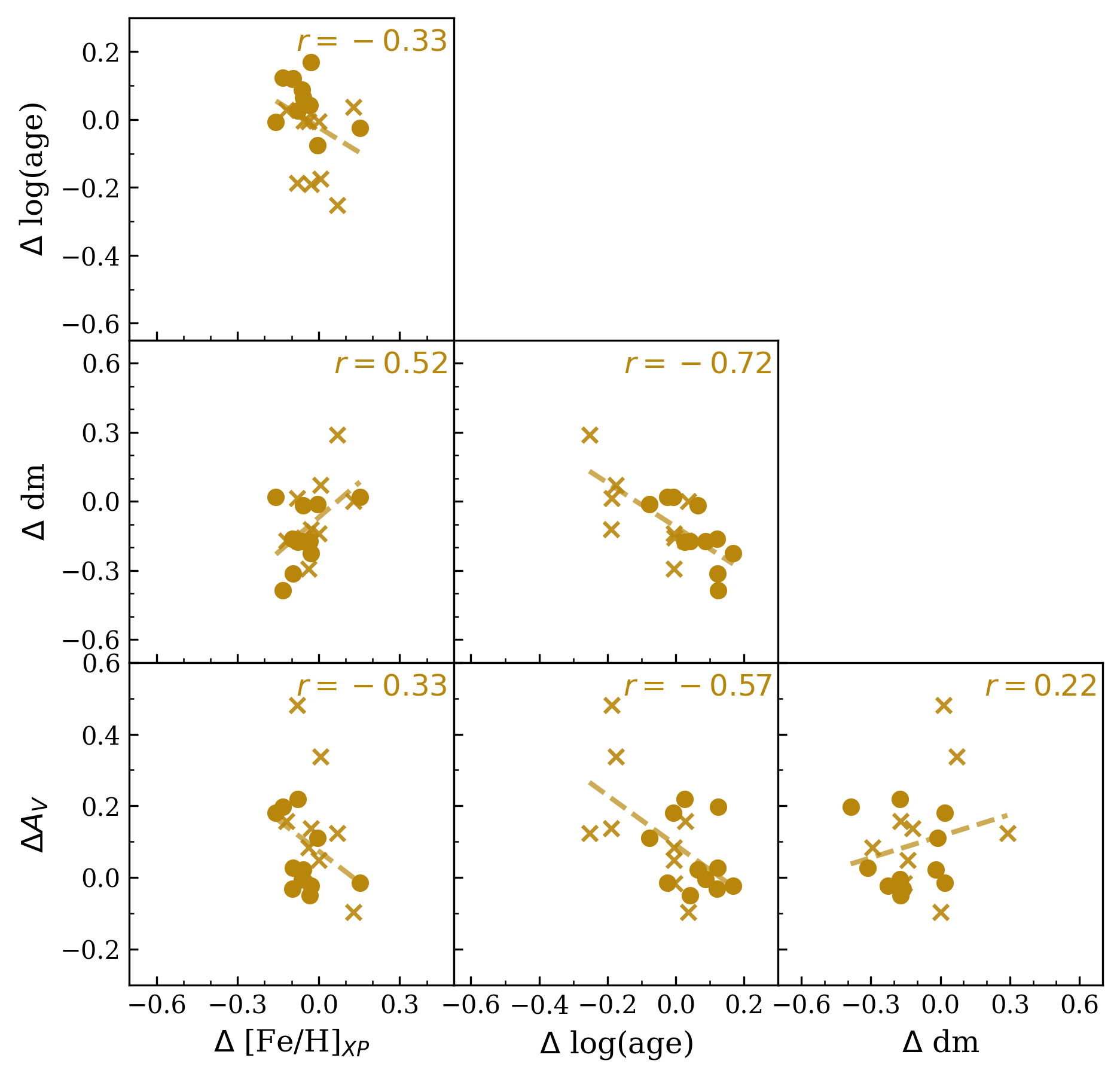}
		\caption{}
		\label{fig:corner_fs}
	\end{subfigure}
	\caption{Corner plot of parameter offsets relative to B19 for the solutions performed (a) without a [Fe/H] prior and (b) with a [Fe/H] prior calculated from the FS catalogue. Two clusters with exceptionally large $\Delta$\logage\ are marked with green markers and are excluded from the correlation calculation in the plot (a). The dashed lines show linear fits to the data, and the Pearson correlation coefficient $r$ is shown for each pair of parameters. Clusters with unpopulated RGB are shown with an X marker.}
	\label{fig:corner_nofix_fs}
\end{figure*}

\subsection{Correlation analysis}
In the previous section we evaluated the accuracy of the parameters independently of each other. However, since isochrone fitting is a multi-parameter fit, we analysed the correlations between the offsets in all fitted parameters.

For the solutions without a metallicity prior (Fig.~\ref{fig:corner_nofix}), we find a strong negative correlation between $\Delta$\logage\ and $\Delta$dm (Pearson correlation coefficient $r=-0.92$, $p < 0.001$). This is to be expected, since increasing the age shifts the MSTO lower and decreasing the distance modulus shifts the isochrone upwards in the CMD, thus compensating each other. This degeneracy has also been noted in other isochrone-fitting works (e.g. \citet{Ying2023}). Two clusters with no RGB (NGC 1912 and NGC 6475, marked as green) stand out from this clear trend by having a large $\Delta$\logage\ and were not included in the correlation calculations. These clusters also show significant offsets in the other two parameters, and, in case of NGC 6475, also $\Delta\mathrm{[Fe/H]_{fit}}$. Although their isochrones appear consistent with the photometric data upon visual inspection (Figure \ref{fig:all-cmds}), the absence of an RGB substantially reduces the reliability of the fits.

The metallicity offset $\Delta\mathrm{[Fe/H]_{fit}}$ with respect to the B19 values also shows noticeable correlations with all other parameters: with \logage\ ($r=-0.66, p = 0.003$), dm ($r=+0.71, p = 0.001$), and $A_V$ ($r=-0.73, p < 0.001$). An isochrone with a higher [Fe/H] will have a redder main sequence and RGB, which can be compensated by increasing $A_V$, lowering the turn-off age, or shifting the dm. As mentioned in Section \ref{sect:dm-Av}, $A_V$ values in this regime exhibit less scatter compared to the fits with Gaia XP prior. In the absence of a [Fe/H] constraint, the fit is free to adjust the metallicity jointly with $A_V$, so the [Fe/H] estimate can absorb the uncertainties. Whereas when [Fe/H] prior is provided, the degeneracy forces this tension into $A_V$, increasing its dispersion.  

For the solutions with a [Fe/H] prior from the FS catalogue (Fig.~\ref{fig:corner_fs}), the $\Delta$\logage--$\Delta$dm correlation persists, but with a weaker slope ($r=-0.72, p < 0.001$), while other strong correlations are weakened. The extinction offsets are also reduced and remain close to those of the high-resolution spectroscopy solutions, reflecting the small systematic offset in [Fe/H] with respect to B19.  

The ARC catalogue (Figure \ref{fig:corner_arc}) shows a correlation of  $r=-0.55$ ($p = 0.11$) between $\Delta$[Fe/H]$_{XP}$ and $\Delta A_V$, in contrast to $r=-0.33$ ($p = 0.16$) for the FS catalogue. This is consistent with ARC's larger median metallicity offset ($-0.1$ dex relative to FS, see Table \ref{tab:age-compare}), which in turn propagates into larger extinction offsets. The ZGR catalogue shows similar behaviour but with slightly weaker correlations ($r=-0.50, p = 0.24$ between $\Delta$[Fe/H]$_{XP}$ and $\Delta A_V$), see Figure \ref{fig:corner_zgr}.

The fits that used the HRS [Fe/H] prior (Figure \ref{fig:corner_fix}) do not show a correlation between $\Delta$\logage\ and $\Delta$dm, but instead some correlation between $\Delta$\logage\ and $\Delta A_V$ ($r=-0.60, p = 0.005$). However, the clusters driving these offsets are those lacking a well-populated RGB, whereas for other clusters these offsets are not correlated.

These correlations demonstrate the presence of \logage-dm degeneracy in isochrone fitting, as well as highlight that [Fe/H] errors are a primary source of extinction biases. The strength of these correlations varies between the adopted [Fe/H] priors: the ARC prior showed the strongest metallicity–extinction coupling, whereas the FS was closest to the benchmark values of B19. Adopting these priors helps diminishing the strong correlations between all four parameters seen in the regime with no [Fe/H] prior at all.

\subsection{Comparison of cluster parameters with NN studies}
\begin{figure*}[ht!]
	\centering
	\includegraphics[width=1\linewidth]{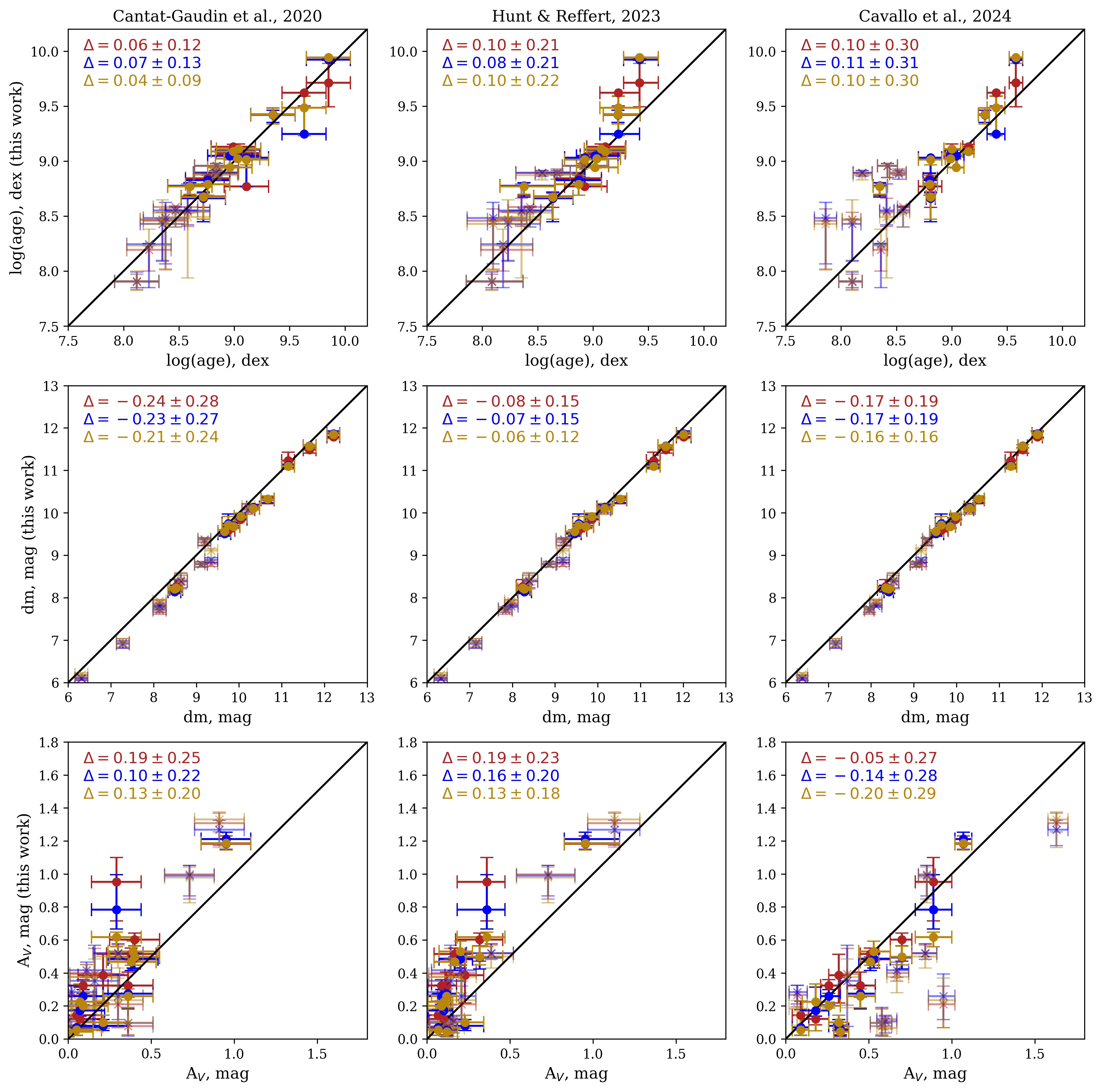}
	\caption{Parameter estimates for age, distance modulus and extinction (rows, Y axis) of the OCs determined with a [Fe/H] prior based on Gaia XP metallicity compared to those estimates reported by the studies of CG20, HR23, and \citet{cavallo24} on the X axis. The same colour-coding scheme applies as in Fig. \ref{fig:age-compare}. The legend in the upper corner of the figures shows the median absolute difference and the RMSD of the respective parameter. The OCs with three or fewer RGB stars are marked with an X symbol.}
	\label{fig:age-compare-ml}
\end{figure*}

To further evaluate the impact of incorporating a metallicity prior ([Fe/H]) in our parameter estimates, we compared our results with three recent studies that derive cluster parameters using trained neural networks: CG20, HR23, and \citet{cavallo24}. These studies did not rely on spectroscopic information for individual clusters but instead obtained parameters directly from their machine-learning approaches, bypassing the need for standard isochrone fitting procedures. We demonstrate these results in Figure \ref{fig:age-compare-ml}, in a similar fashion: rows show the three parameters obtained in the fit, whereas columns denote the study to which the comparison was made.

CG20 (left column) employed a neural network trained on 347 reference clusters, 269 of which were also part of the B19 sample. They do not derive or fix [Fe/H] as one of the input parameters, instead their network was trained on a grid that covered a large range of metallicities. The agreement in the cluster ages between CG20 and our work is good, with a comparable level of accuracy to the agreement we observed with B19. This is expected due to the overlap in training data used by their network.

However, the distance modulus (dm) values derived by CG20 show a significant negative systematic offset and dispersion compared to our results. The extinction values also exhibit a large offset and RMSD, indicating discrepancies in their estimation for clusters of not just high, but also low extinction.

HR23 (middle column) included metallicity information in their analysis in the same implicit way as CG20. The cluster ages derived in their work show poorer agreement with ours with a dispersion of $\sim$0.2 dex. This highlights the limitations of their approach for precise age determinations.

In contrast, the dm values from HR23 align well with our results and even show better agreement compared to the B19 sample. However, similar to CG20, the extinction values $A_{V}$ follow a similar pattern of significant offsets and large dispersion.

As previously discussed, \citet{cavallo24} (left column) utilised neural networks to estimate cluster parameters. Their age estimates, particularly for young clusters and those lacking a significant RGB population, show large discrepancies and a high level of scatter. This issue is consistent with the challenges highlighted in the CG20 and HR23 analyses. The overestimation of ages was also noted by \citet{cavallo24} during the scientific verification of their whole dataset.

The distance modulus results from \citet{cavallo24} are similar in accuracy to those derived from B19. The extinction values $A_{V}$ show considerable scatter, especially for young clusters with no RGB population, reinforcing the limitations of their methodology in handling such cases.

In summary, these comparisons underscore the importance of incorporating reliable metallicity priors, such as those derived from high-resolution spectroscopy or Gaia XP, to enhance the accuracy and consistency of cluster parameter estimates. While neural network-based approaches offer promising alternatives considering the demands of large datasets and an ever-growing number of clusters, they face significant challenges, particularly in estimating ages, metallicity and extinction values for specific types of clusters. As a potential solution, one can use the [Fe/H] values based on Gaia XP spectra as priors in the NN.

\section{Discussion}\label{sect:discussion}
In this study we relied exclusively on Gaia DR3 data, including Gaia photometry, parallax, and studies that use Gaia XP spectra, for deriving cluster metallicities and use them in age determination through isochrone fitting. While Gaia data provides extensive coverage and unprecedented astrometric precision, the low resolution of the XP spectra introduces several challenges in accurately estimating stellar parameters, particularly in metallicities. Machine learning techniques allow us to extract metallicity information from these XP spectra, but they are in turn limited by the parameter range of the training data.

The [M/H] that we obtained from the ARC catalogue shows a far larger negative offset compared to the values derived from high-resolution spectroscopy in contrast with the ZGR and FS catalogues. The algorithm of ARC was undertrained for hot stars, since the reliable $T_{\mathrm{eff}}$ range of the training data is 3107 K to 6867 K, which corresponds to the unreddened ${BP} - {RP}$ colour indices between 0.5 and 2.3. The imposed cut of ${BP} - {RP} > 0.5$ mag does eliminate most of these hot stars that show lower metallicities than the rest; however, that is not the case for several clusters in the sample. In the clusters with higher $A_V$, the observed ${BP} - {RP}$ value is affected by the extinction, which in turn makes the hot stars appear redder and pass the cut, thereby lowering the derived [Fe/H] and increasing the corresponding error (see IC 2714 in Fig. \ref{fig:bprp_all_1}).

We tested a stricter colour range of $0.8<BP-RP<1.2$ to see if it would improve the metallicity estimates. However, the median improvement for ARC was $\sim$0.04 dex, which is negligible within the catalogue uncertainty. Moreover, this cut drastically decreased the amount of usable members, which in turn cut in half our sample of clusters with >100 members in each catalogue.

We performed another test in which we first corrected the ${BP} - {RP}$ values using the $A_V$ values determined in B19 and the conversion (\ref{eq:av-conversion}), then we repeated the process of metallicity determination. The $A_V$ correction did only improve both the median difference and RMSD of ARC [Fe/H] by $\sim$0.02 dex, which is negligible and does not subsequently lead to significant differences in cluster ages, as seen in the results obtained with three XP-based metallicity catalogues. This demonstrates that our initial approach allows us to derive reliable cluster parameters using only Gaia data, in contrast to B19, which based the $A_V$ prior on DAML and MWSC catalogues.

Another reason for the random error of [Fe/H] measurements is the main population of the open clusters in the sample, which is main sequence stars. \citet{andrae2023} report an error of $\sim$0.33 dex at $G \sim 17$, which is caused by the noise in the XP spectra that have a lower S/N ratio with increasing magnitude.

The FS catalogue was also trained on the APOGEE data, much like ARC, yet it consistently shows better agreement with the results of B19 both in derived metallicities and cluster parameters. While ARC and ZGR predict [M/H], FS's methodology predicts [Fe/H] along with [C/Fe], [N/Fe] and [$\alpha$/M] and with a better precision, as it uses uncertainty estimation and additionally uses 2MASS photometry to break temperature degeneracies. In this study we equated [M/H] to [Fe/H] for these catalogues, which led to noticeable discrepancies in the results.

Between the derived cluster parameters, the largest discrepancies with literature values were seen in $A_V$, which could be caused by a combination of factors. First, in our fitting procedure with ASteCA we include unresolved binaries, which can mimic the effects of extinction by making stars appear redder, if not accounted for. Meanwhile, the BASE-9 code used by B19 did not include binary systems. Second, the majority of clusters in our sample have relatively low extinction values ($A_V$ $< 0.5$), where even small differences in photometry or assumptions can lead to large relative deviations from other studies. Third, the largest discrepancies between our $A_V$ estimates and those of B19 are found at $A_V > 0.5$ and persist across all solutions, independent of whether they used a Gaia XP prior, an HRS prior, or no metallicity prior. These extinction estimates lie outside the uncertainty threshold of \(\sigma_{A_V} = \frac{1}{3} A_V\) adopted by B19, which was not applied in our analysis. Introducing an extinction prior from an independent source and marginalising over it would help to reduce these discrepancies. Finally, some clusters lack a red giant branch or red clump, which typically provide strong constraints on extinction in CMD fitting. In their absence, extinction estimates become more sensitive to the placement and shape of the main sequence,

Lastly, we imposed a floor of 50\% for the membership probability of the cluster members to be included in the analysis. This allowed us to include the most amount of stars for [Fe/H] determination. However, this cut in turn included individual stars or groups of stars that did not resemble cluster members upon a visual inspection of the CMD (see NGC 3532 in Fig. \ref{fig:all-cmds}), but we did not exclude them to preserve the consistency of the methodology. Since ASteCA does not include assigning weights to the stars depending on their MP, their presence could potentially affect the goodness of the fit and the fit parameters.

\section{Conclusion} \label{sect:conclusions}
We conducted an analysis of 20 Milky Way open clusters, focusing on their metallicity and cluster parameter determination using data from the Gaia mission alone. We used three existing catalogues that determined stellar parameters from all available 220 million Gaia XP spectra, from which we derived the [Fe/H] values of open clusters, observing differences of 0.1 dex for the ZGR catalogue, 0.15 dex for the ARC catalogue and 0.07 dex for the FS catalogue when matched against high-resolution spectroscopy values. The median errors in [Fe/H] measurements per cluster using these catalogues were found to be 0.12 dex, 0.13 dex, and 0.07 dex, respectively. These findings indicate that while the Gaia XP spectra yield reliable metallicity estimates, they exhibit systematic differences from traditional methods, particularly at lower metallicities. Moreover, the better agreement of FS metallicities was achieved by imposing a stricter cut on the accepted metallicity values in the clusters of the study.

By utilising the ASteCA code, we estimated the parameters of open clusters by comparing synthetic clusters generated with PARSEC isochrones. Our results show a good agreement with the values of B19. This method allowed us to determine cluster ages with the comparable accuracy to their study that required costly and time-consuming spectroscopic observations to lift the age-metallicity degeneracy. Our approach effectively reduced uncertainties and degeneracies between the parameters by incorporating [Fe/H] as a prior, enhancing the accuracy of our age determinations primarily for open clusters without a populated RGB. The results indicate an improvement in age determination accuracy compared to the ages determined with a neural network by CG20, HR23 and \citet{cavallo24}. However, the estimates of extinction and distance modulus do not demonstrate a significant uncertainty reduction. 

Our findings demonstrate the potential of utilising only Gaia data for studying the parameters of Milky Way open clusters, after applying sufficient cuts to the Gaia XP metallicity estimates, thereby enhancing our understanding of stellar evolution within these groupings. Looking forwards, this method can be expanded to the whole sky to encompass all clusters. Moreover, the estimations of $[\alpha / \text{Fe}]$ from Gaia XP spectra \citep{rvs-cnn} can help in further constraints of the chemical composition of the open clusters and its influence of the cluster parameters and the evolution of the galaxy. While these methods remain model-dependent, the integration of asteroseismology and chemical clocks to determine the ages will further refine the methodologies of cluster age determination.

\section{Data availability}
Tables \ref{appendix_params_table1} and \ref{appendix_params_table2} with the derived cluster parameters are only available in electronic form at the CDS via anonymous ftp to \url{https://cdsarc.cds.unistra.fr/} (130.79.128.5) or via \url{http://cdsweb.u-strasbg.fr/cgi-bin/qcat?J/A+A/}.

\begin{acknowledgements}
The authors thank the anonymous referee for their suggestions, which helped improve the quality of this manuscript. The authors also thank G. Perren for his extensive advice regarding the applications of ASteCA. 
This work used the following open source software: TOPCAT \citep{topcat}, ezpadova. This work has made use of data from the European Space Agency (ESA) mission {\it Gaia} (\url{https://www.cosmos.esa.int/gaia}), processed by the {\it Gaia} Data Processing and Analysis Consortium (DPAC, \url{https://www.cosmos.esa.int/web/gaia/dpac/consortium}). Funding for the DPAC has been provided by national institutions, in particular the institutions participating in the {\it Gaia} Multilateral Agreement. This research has made use of the tool provided by Gaia DPAC (\url{https://www.cosmos.esa.int/web/gaia/dr3-software-tools}) to reproduce (E)DR3 Gaia photometric uncertainties described in the GAIA-C5-TN-UB-JMC-031 technical note using data in \citet{riello2021}. This research or product makes use of public auxiliary data provided by ESA/Gaia/DPAC/CU5 and prepared by Carine Babusiaux. This research has made use of NASA's Astrophysics Data System Bibliographic Services. This work was supported by NOVA (the Netherlands Research School for Astronomy).
\end{acknowledgements}
\bibliographystyle{aa}
\bibliography{oc-gaia.bib}

\begin{thebibliography}{47}
\expandafter\ifx\csname natexlab\endcsname\relax\def\natexlab#1{#1}\fi

\bibitem[{{Abdurro'uf} {et~al.}(2022){Abdurro'uf}, {Accetta}, {Aerts}, {Silva
  Aguirre}, {Ahumada}, {Ajgaonkar}, {Filiz Ak}, {Alam}, {Allende Prieto},
  {Almeida}, {Anders}, {Anderson}, {Andrews}, {Anguiano}, {Aquino-Ort{\'\i}z},
  {Arag{\'o}n-Salamanca}, {Argudo-Fern{\'a}ndez}, {Ata}, {Aubert},
  {Avila-Reese}, {Badenes}, {Barb{\'a}}, {Barger}, {Barrera-Ballesteros},
  {Beaton}, {Beers}, {Belfiore}, {Bender}, {Bernardi}, {Bershady}, {Beutler},
  {Bidin}, {Bird}, {Bizyaev}, {Blanc}, {Blanton}, {Boardman}, {Bolton},
  {Boquien}, {Borissova}, {Bovy}, {Brandt}, {Brown}, {Brownstein}, {Brusa},
  {Buchner}, {Bundy}, {Burchett}, {Bureau}, {Burgasser}, {Cabang}, {Campbell},
  {Cappellari}, {Carlberg}, {Wanderley}, {Carrera}, {Cash}, {Chen}, {Chen},
  {Cherinka}, {Chiappini}, {Choi}, {Chojnowski}, {Chung}, {Clerc}, {Cohen},
  {Comerford}, {Comparat}, {da Costa}, {Covey}, {Crane}, {Cruz-Gonzalez},
  {Culhane}, {Cunha}, {Dai}, {Damke}, {Darling}, {Davidson}, {Davies},
  {Dawson}, {De Lee}, {Diamond-Stanic}, {Cano-D{\'\i}az}, {S{\'a}nchez},
  {Donor}, {Duckworth}, {Dwelly}, {Eisenstein}, {Elsworth}, {Emsellem},
  {Eracleous}, {Escoffier}, {Fan}, {Farr}, {Feng}, {Fern{\'a}ndez-Trincado},
  {Feuillet}, {Filipp}, {Fillingham}, {Frinchaboy}, {Fromenteau}, {Galbany},
  {Garc{\'\i}a}, {Garc{\'\i}a-Hern{\'a}ndez}, {Ge}, {Geisler}, {Gelfand},
  {G{\'e}ron}, {Gibson}, {Goddy}, {Godoy-Rivera}, {Grabowski}, {Green},
  {Greener}, {Grier}, {Griffith}, {Guo}, {Guy}, {Hadjara}, {Harding},
  {Hasselquist}, {Hayes}, {Hearty}, {Hern{\'a}ndez}, {Hill}, {Hogg},
  {Holtzman}, {Horta}, {Hsieh}, {Hsu}, {Hsu}, {Huber}, {Huertas-Company},
  {Hutchinson}, {Hwang}, {Ibarra-Medel}, {Chitham}, {Ilha}, {Imig}, {Jaekle},
  {Jayasinghe}, {Ji}, {Johnson}, {Jones}, {J{\"o}nsson}, {Katkov}, {Khalatyan},
  {Kinemuchi}, {Kisku}, {Knapen}, {Kneib}, {Kollmeier}, {Kong}, {Kounkel},
  {Kreckel}, {Krishnarao}, {Lacerna}, {Lane}, {Langgin}, {Lavender}, {Law},
  {Lazarz}, {Leung}, {Leung}, {Lewis}, {Li}, {Li}, {Lian}, {Liang}, {Lin},
  {Lin}, {Lin}, {Lintott}, {Long}, {Longa-Pe{\~n}a}, {L{\'o}pez-Cob{\'a}},
  {Lu}, {Lundgren}, {Luo}, {Mackereth}, {de la Macorra}, {Mahadevan},
  {Majewski}, {Manchado}, {Mandeville}, {Maraston}, {Margalef-Bentabol},
  {Masseron}, {Masters}, {Mathur}, {McDermid}, {Mckay}, {Merloni},
  {Merrifield}, {Meszaros}, {Miglio}, {Di Mille}, {Minniti}, {Minsley},
  {Monachesi}, {Moon}, {Mosser}, {Mulchaey}, {Muna}, {Mu{\~n}oz}, {Myers},
  {Myers}, {Nadathur}, {Nair}, {Nandra}, {Neumann}, {Newman}, {Nidever},
  {Nikakhtar}, {Nitschelm}, {O'Connell}, {Garma-Oehmichen}, {Luan Souza de
  Oliveira}, {Olney}, {Oravetz}, {Ortigoza-Urdaneta}, {Osorio}, {Otter},
  {Pace}, {Padilla}, {Pan}, {Pan}, {Parikh}, {Parker}, {Peirani}, {Pe{\~n}a
  Ram{\'\i}rez}, {Penny}, {Percival}, {Perez-Fournon}, {Pinsonneault},
  {Poidevin}, {Poovelil}, {Price-Whelan}, {B{\'a}rbara de Andrade Queiroz},
  {Raddick}, {Ray}, {Rembold}, {Riddle}, {Riffel}, {Riffel}, {Rix}, {Robin},
  {Rodr{\'\i}guez-Puebla}, {Roman-Lopes}, {Rom{\'a}n-Z{\'u}{\~n}iga}, {Rose},
  {Ross}, {Rossi}, {Rubin}, {Salvato}, {S{\'a}nchez}, {S{\'a}nchez-Gallego},
  {Sanderson}, {Santana Rojas}, {Sarceno}, {Sarmiento}, {Sayres}, {Sazonova},
  {Schaefer}, {Schiavon}, {Schlegel}, {Schneider}, {Schultheis}, {Schwope},
  {Serenelli}, {Serna}, {Shao}, {Shapiro}, {Sharma}, {Shen}, {Shetrone}, {Shu},
  {Simon}, {Skrutskie}, {Smethurst}, {Smith}, {Sobeck}, {Spoo}, {Sprague},
  {Stark}, {Stassun}, {Steinmetz}, {Stello}, {Stone-Martinez},
  {Storchi-Bergmann}, {Stringfellow}, {Stutz}, {Su}, {Taghizadeh-Popp},
  {Talbot}, {Tayar}, {Telles}, {Teske}, {Thakar}, {Theissen}, {Tkachenko},
  {Thomas}, {Tojeiro}, {Hernandez Toledo}, {Troup}, {Trump}, {Trussler},
  {Turner}, {Tuttle}, {Unda-Sanzana}, {V{\'a}zquez-Mata}, {Valentini},
  {Valenzuela}, {Vargas-Gonz{\'a}lez}, {Vargas-Maga{\~n}a}, {Alfaro},
  {Villanova}, {Vincenzo}, {Wake}, {Warfield}, {Washington}, {Weaver},
  {Weijmans}, {Weinberg}, {Weiss}, {Westfall}, {Wild}, {Wilde}, {Wilson},
  {Wilson}, {Wilson}, {Wolf}, {Wood-Vasey}, {Yan}, {Zamora}, {Zasowski},
  {Zhang}, {Zhao}, {Zheng}, {Zheng}, \& {Zhu}}]{apogee}
{Abdurro'uf}, {Accetta}, K., {Aerts}, C., {et~al.} 2022, \apjs, 259, 35

\bibitem[{{Andrae} {et~al.}(2023{\natexlab{a}}){Andrae}, {Fouesneau}, {Sordo},
  {Bailer-Jones}, {Dharmawardena}, {Rybizki}, {De Angeli}, {Lindstr{\o}m},
  {Marshall}, {Drimmel}, {Korn}, {Soubiran}, {Brouillet}, {Casamiquela}, {Rix},
  {Abreu Aramburu}, {{\'A}lvarez}, {Bakker}, {Bellas-Velidis}, {Bijaoui},
  {Brugaletta}, {Burlacu}, {Carballo}, {Chaoul}, {Chiavassa}, {Contursi},
  {Cooper}, {Creevey}, {Dafonte}, {Dapergolas}, {de Laverny}, {Delchambre},
  {Demouchy}, {Edvardsson}, {Fr{\'e}mat}, {Garabato}, {Garc{\'\i}a-Lario},
  {Garc{\'\i}a-Torres}, {Gavel}, {Gomez}, {Gonz{\'a}lez-Santamar{\'\i}a},
  {Hatzidimitriou}, {Heiter}, {Jean-Antoine Piccolo}, {Kontizas}, {Kordopatis},
  {Lanzafame}, {Lebreton}, {Licata}, {Livanou}, {Lobel}, {Lorca}, {Magdaleno
  Romeo}, {Manteiga}, {Marocco}, {Mary}, {Nicolas}, {Ordenovic}, {Pailler},
  {Palicio}, {Pallas-Quintela}, {Panem}, {Pichon}, {Poggio}, {Recio-Blanco},
  {Riclet}, {Robin}, {Santove{\~n}a}, {Sarro}, {Schultheis}, {Segol},
  {Silvelo}, {Slezak}, {Smart}, {S{\"u}veges}, {Th{\'e}venin}, {Torralba
  Elipe}, {Ulla}, {Utrilla}, {Vallenari}, {van Dillen}, {Zhao}, \&
  {Zorec}}]{gsp-phot}
{Andrae}, R., {Fouesneau}, M., {Sordo}, R., {et~al.} 2023{\natexlab{a}}, \aap,
  674, A27

\bibitem[{{Andrae} {et~al.}(2023{\natexlab{b}}){Andrae}, {Rix}, \&
  {Chandra}}]{andrae2023}
{Andrae}, R., {Rix}, H.-W., \& {Chandra}, V. 2023{\natexlab{b}}, \apjs, 267, 8

\bibitem[{{Angus} {et~al.}(2019){Angus}, {Morton}, {Foreman-Mackey}, {van
  Saders}, {Curtis}, {Kane}, {Bedell}, {Kiman}, {Hogg}, \&
  {Brewer}}]{angus2019}
{Angus}, R., {Morton}, T.~D., {Foreman-Mackey}, D., {et~al.} 2019, \aj, 158,
  173

\bibitem[{{Bianchini} \& {Mastrobuono-Battisti}(2024)}]{bianchini2024}
{Bianchini}, P. \& {Mastrobuono-Battisti}, A. 2024, \mnras, 527, L32

\bibitem[{{Bossini} {et~al.}(2019){Bossini}, {Vallenari}, {Bragaglia},
  {Cantat-Gaudin}, {Sordo}, {Balaguer-N{\'u}{\~n}ez}, {Jordi}, {Moitinho},
  {Soubiran}, {Casamiquela}, {Carrera}, \& {Heiter}}]{bossini2019}
{Bossini}, D., {Vallenari}, A., {Bragaglia}, A., {et~al.} 2019, \aap, 623, A108

\bibitem[{{Cantat-Gaudin} {et~al.}(2020){Cantat-Gaudin}, {Anders},
  {Castro-Ginard}, {Jordi}, {Romero-G{\'o}mez}, {Soubiran}, {Casamiquela},
  {Tarricq}, {Moitinho}, {Vallenari}, {Bragaglia}, {Krone-Martins}, \&
  {Kounkel}}]{cantat2020}
{Cantat-Gaudin}, T., {Anders}, F., {Castro-Ginard}, A., {et~al.} 2020, \aap,
  640, A1

\bibitem[{Cantat-Gaudin {et~al.}(2018)Cantat-Gaudin, Jordi, Vallenari,
  Bragaglia, Balaguer-Núñez, Bossini, Casamiquela, Soubiran, Moitinho,
  Castro-Ginard, \& Krone-Martins}]{cantat2018}
Cantat-Gaudin, T., Jordi, C., Vallenari, A., {et~al.} 2018, \aap, 618, A93

\bibitem[{Castro-Ginard {et~al.}(2019)Castro-Ginard, Jordi, Luri,
  Cantat-Gaudin, Balaguer-Núñez, \& Soubiran}]{castro2019}
Castro-Ginard, A., Jordi, C., Luri, X., {et~al.} 2019, \aap, 627, A35

\bibitem[{{Cavallo} {et~al.}(2024){Cavallo}, {Spina}, {Carraro}, {Magrini},
  {Poggio}, {Cantat-Gaudin}, {Pasquato}, {Lucatello}, {Ortolani}, \&
  {Schiappacasse-Ulloa}}]{cavallo24}
{Cavallo}, L., {Spina}, L., {Carraro}, G., {et~al.} 2024, \aj, 167, 12

\bibitem[{{Dias} {et~al.}(2002){Dias}, {Alessi}, {Moitinho}, \&
  {L{\'e}pine}}]{dias2002}
{Dias}, W.~S., {Alessi}, B.~S., {Moitinho}, A., \& {L{\'e}pine}, J.~R.~D. 2002,
  \aap, 389, 871

\bibitem[{{Dias} {et~al.}(2021){Dias}, {Monteiro}, {Moitinho}, {L{\'e}pine},
  {Carraro}, {Paunzen}, {Alessi}, \& {Villela}}]{dias2021}
{Dias}, W.~S., {Monteiro}, H., {Moitinho}, A., {et~al.} 2021, \mnras, 504, 356

\bibitem[{{Fallows} \& {Sanders}(2024)}]{fs24}
{Fallows}, C.~P. \& {Sanders}, J.~L. 2024, \mnras, 531, 2126

\bibitem[{{Gaia Collaboration} {et~al.}(2018){Gaia Collaboration}, Brown,
  Vallenari, Prusti, de~Bruijne, Babusiaux, Bailer-Jones, Biermann, Evans,
  Eyer, Jansen, Klioner, Lammers, Lindegren, Luri, Mignard, Panem, Pourbaix,
  Randich, Sartoretti, Siddiqui, Soubiran, \& van Leeuwen}]{gaia2018}
{Gaia Collaboration}, Brown, A. G.~A., Vallenari, A., {et~al.} 2018, \aap, 616,
  A1

\bibitem[{{Gaia Collaboration} {et~al.}(2023){Gaia Collaboration}, {Creevey},
  {Sarro}, {Lobel}, {Pancino}, {Andrae}, {Smart}, {Clementini}, {Heiter},
  {Korn}, {Fouesneau}, {Fr{\'e}mat}, {De Angeli}, {Vallenari}, {Harrison},
  {Th{\'e}venin}, {Reyl{\'e}}, {Sordo}, {Garofalo}, {Brown}, {Eyer}, {Prusti},
  {de Bruijne}, {Arenou}, {Babusiaux}, {Biermann}, {Ducourant}, {Evans},
  {Guerra}, {Hutton}, {Jordi}, {Klioner}, {Lammers}, {Lindegren}, {Luri},
  {Mignard}, {Panem}, {Pourbaix}, {Randich}, {Sartoretti}, {Soubiran}, {Tanga},
  {Walton}, {Bailer-Jones}, {Bastian}, {Drimmel}, {Jansen}, {Katz}, {Lattanzi},
  {van Leeuwen}, {Bakker}, {Cacciari}, {Casta{\~n}eda}, {Fabricius},
  {Galluccio}, {Guerrier}, {Masana}, {Messineo}, {Mowlavi}, {Nicolas},
  {Nienartowicz}, {Pailler}, {Panuzzo}, {Riclet}, {Roux}, {Seabroke},
  {Gracia-Abril}, {Portell}, {Teyssier}, {Altmann}, {Audard}, {Bellas-Velidis},
  {Benson}, {Berthier}, {Blomme}, {Burgess}, {Busonero}, {Busso},
  {C{\'a}novas}, {Carry}, {Cellino}, {Cheek}, {Damerdji}, {Davidson}, {de
  Teodoro}, {Nu{\~n}ez Campos}, {Delchambre}, {Dell'Oro}, {Esquej},
  {Fern{\'a}ndez-Hern{\'a}ndez}, {Fraile}, {Garabato}, {Garc{\'\i}a-Lario},
  {Gosset}, {Haigron}, {Halbwachs}, {Hambly}, {Hern{\'a}ndez}, {Hestroffer},
  {Hodgkin}, {Holl}, {Jan{\ss}en}, {Jevardat de Fombelle}, {Jordan},
  {Krone-Martins}, {Lanzafame}, {L{\"o}ffler}, {Marchal}, {Marrese},
  {Moitinho}, {Muinonen}, {Osborne}, {Pauwels}, {Recio-Blanco}, {Riello},
  {Rimoldini}, {Roegiers}, {Rybizki}, {Siopis}, {Smith}, {Sozzetti}, {Utrilla},
  {van Leeuwen}, {Abbas}, {{\'A}brah{\'a}m}, {Abreu Aramburu}, {Aerts},
  {Aguado}, {Ajaj}, {Aldea-Montero}, {Altavilla}, {{\'A}lvarez}, {Alves},
  {Anders}, {Anderson}, {Anglada Varela}, {Antoja}, {Baines}, {Baker},
  {Balaguer-N{\'u}{\~n}ez}, {Balbinot}, {Balog}, {Barache}, {Barbato},
  {Barros}, {Barstow}, {Bartolom{\'e}}, {Bassilana}, {Bauchet}, {Becciani},
  {Bellazzini}, {Berihuete}, {Bernet}, {Bertone}, {Bianchi}, {Binnenfeld},
  {Blanco-Cuaresma}, {Boch}, {Bombrun}, {Bossini}, {Bouquillon}, {Bragaglia},
  {Bramante}, {Breedt}, {Bressan}, {Brouillet}, {Brugaletta}, {Bucciarelli},
  {Burlacu}, {Butkevich}, {Buzzi}, {Caffau}, {Cancelliere}, {Cantat-Gaudin},
  {Carballo}, {Carlucci}, {Carnerero}, {Carrasco}, {Casamiquela}, {Castellani},
  {Castro-Ginard}, {Chaoul}, {Charlot}, {Chemin}, {Chiaramida}, {Chiavassa},
  {Chornay}, {Comoretto}, {Contursi}, {Cooper}, {Cornez}, {Cowell}, {Crifo},
  {Cropper}, {Crosta}, {Crowley}, {Dafonte}, {Dapergolas}, {David}, {de
  Laverny}, {De Luise}, {De March}, {De Ridder}, {de Souza}, {de Torres}, {del
  Peloso}, {del Pozo}, {Delbo}, {Delgado}, {Delisle}, {Demouchy},
  {Dharmawardena}, {Di Matteo}, {Diakite}, {Diener}, {Distefano}, {Dolding},
  {Enke}, {Fabre}, {Fabrizio}, {Faigler}, {Fedorets}, {Fernique}, {Figueras},
  {Fournier}, {Fouron}, {Fragkoudi}, {Gai}, {Garcia-Gutierrez},
  {Garcia-Reinaldos}, {Garc{\'\i}a-Torres}, {Gavel}, {Gavras}, {Gerlach},
  {Geyer}, {Giacobbe}, {Gilmore}, {Girona}, {Giuffrida}, {Gomel}, {Gomez},
  {Gonz{\'a}lez-N{\'u}{\~n}ez}, {Gonz{\'a}lez-Santamar{\'\i}a},
  {Gonz{\'a}lez-Vidal}, {Granvik}, {Guillout}, {Guiraud},
  {Guti{\'e}rrez-S{\'a}nchez}, {Guy}, {Hatzidimitriou}, {Hauser}, {Haywood},
  {Helmer}, {Helmi}, {Hilger}, {Sarmiento}, {Hidalgo}, {H{\l}adczuk}, {Hobbs},
  {Holland}, {Huckle}, {Jardine}, {Jasniewicz}, {Jean-Antoine Piccolo},
  {Jim{\'e}nez-Arranz}, {Juaristi Campillo}, {Julbe}, {Karbevska}, {Kervella},
  {Khanna}, {Kordopatis}, {K{\'o}sp{\'a}l}, {Kostrzewa-Rutkowska},
  {Kruszy{\'n}ska}, {Kun}, {Laizeau}, {Lambert}, {Lanza}, {Lasne}, {Le
  Campion}, {Lebreton}, {Lebzelter}, {Leccia}, {Leclerc}, {Lecoeur-Taibi},
  {Liao}, {Licata}, {Lindstr{\o}m}, {Lister}, {Livanou}, {Lorca}, {Loup},
  {Madrero Pardo}, {Magdaleno Romeo}, {Managau}, {Mann}, {Manteiga},
  {Marchant}, {Marconi}, {Marcos}, {Marcos Santos}, {Mar{\'\i}n Pina},
  {Marinoni}, {Marocco}, {Marshall}, {Martin Polo}, {Mart{\'\i}n-Fleitas},
  {Marton}, {Mary}, {Masip}, {Massari}, {Mastrobuono-Battisti}, {Mazeh},
  {McMillan}, {Messina}, {Michalik}, {Millar}, {Mints}, {Molina}, {Molinaro},
  {Moln{\'a}r}, {Monari}, {Mongui{\'o}}, {Montegriffo}, {Montero}, {Mor},
  {Mora}, {Morbidelli}, {Morel}, {Morris}, {Muraveva}, {Murphy}, {Musella},
  {Nagy}, {Noval}, {Oca{\~n}a}, {Ogden}, {Ordenovic}, {Osinde}, {Pagani},
  {Pagano}, {Palaversa}, {Palicio}, {Pallas-Quintela}, {Panahi},
  {Payne-Wardenaar}, {Pe{\~n}alosa Esteller}, {Penttil{\"a}}, {Pichon},
  {Piersimoni}, {Pineau}, {Plachy}, {Plum}, {Poggio}, {Pr{\v{s}}a}, {Pulone},
  {Racero}, {Ragaini}, {Rainer}, {Raiteri}, {Ramos}, {Ramos-Lerate}, {Re
  Fiorentin}, {Regibo}, {Richards}, {Rios Diaz}, {Ripepi}, {Riva}, {Rix},
  {Rixon}, {Robichon}, {Robin}, {Robin}, {Roelens}, {Rogues}, {Rohrbasser},
  {Romero-G{\'o}mez}, {Rowell}, {Royer}, {Ruz Mieres}, {Rybicki}, {Sadowski},
  {S{\'a}ez N{\'u}{\~n}ez}, {Sagrist{\`a} Sell{\'e}s}, {Sahlmann}, {Salguero},
  {Samaras}, {Sanchez Gimenez}, {Sanna}, {Santove{\~n}a}, {Sarasso},
  {Schultheis}, {Sciacca}, {Segol}, {Segovia}, {S{\'e}gransan}, {Semeux},
  {Shahaf}, {Siddiqui}, {Siebert}, {Siltala}, {Silvelo}, {Slezak}, {Slezak},
  {Snaith}, {Solano}, {Solitro}, {Souami}, {Souchay}, {Spagna}, {Spina},
  {Spoto}, {Steele}, {Steidelm{\"u}ller}, {Stephenson}, {S{\"u}veges},
  {Surdej}, {Szabados}, {Szegedi-Elek}, {Taris}, {Taylor}, {Teixeira},
  {Tolomei}, {Tonello}, {Torra}, {Torra}, {Torralba Elipe}, {Trabucchi},
  {Tsounis}, {Turon}, {Ulla}, {Unger}, {Vaillant}, {van Dillen}, {van Reeven},
  {Vanel}, {Vecchiato}, {Viala}, {Vicente}, {Voutsinas}, {Weiler}, {Wevers},
  {Wyrzykowski}, {Yoldas}, {Yvard}, {Zhao}, {Zorec}, {Zucker}, \&
  {Zwitter}}]{gaiadr3}
{Gaia Collaboration}, {Creevey}, O.~L., {Sarro}, L.~M., {et~al.} 2023, \aap,
  674, A39

\bibitem[{{Gaia Collaboration} {et~al.}(2016){Gaia Collaboration}, {Prusti},
  {de Bruijne}, {Brown}, {Vallenari}, {Babusiaux}, {Bailer-Jones}, {Bastian},
  {Biermann}, {Evans}, {Eyer}, {Jansen}, {Jordi}, {Klioner}, {Lammers},
  {Lindegren}, {Luri}, {Mignard}, {Milligan}, {Panem}, {Poinsignon},
  {Pourbaix}, {Randich}, {Sarri}, {Sartoretti}, {Siddiqui}, {Soubiran},
  {Valette}, {van Leeuwen}, {Walton}, {Aerts}, {Arenou}, {Cropper}, {Drimmel},
  {H{\o}g}, {Katz}, {Lattanzi}, {O'Mullane}, {Grebel}, {Holland}, {Huc},
  {Passot}, {Bramante}, {Cacciari}, {Casta{\~n}eda}, {Chaoul}, {Cheek}, {De
  Angeli}, {Fabricius}, {Guerra}, {Hern{\'a}ndez}, {Jean-Antoine-Piccolo},
  {Masana}, {Messineo}, {Mowlavi}, {Nienartowicz}, {Ord{\'o}{\~n}ez-Blanco},
  {Panuzzo}, {Portell}, {Richards}, {Riello}, {Seabroke}, {Tanga},
  {Th{\'e}venin}, {Torra}, {Els}, {Gracia-Abril}, {Comoretto},
  {Garcia-Reinaldos}, {Lock}, {Mercier}, {Altmann}, {Andrae}, {Astraatmadja},
  {Bellas-Velidis}, {Benson}, {Berthier}, {Blomme}, {Busso}, {Carry},
  {Cellino}, {Clementini}, {Cowell}, {Creevey}, {Cuypers}, {Davidson}, {De
  Ridder}, {de Torres}, {Delchambre}, {Dell'Oro}, {Ducourant}, {Fr{\'e}mat},
  {Garc{\'\i}a-Torres}, {Gosset}, {Halbwachs}, {Hambly}, {Harrison}, {Hauser},
  {Hestroffer}, {Hodgkin}, {Huckle}, {Hutton}, {Jasniewicz}, {Jordan},
  {Kontizas}, {Korn}, {Lanzafame}, {Manteiga}, {Moitinho}, {Muinonen},
  {Osinde}, {Pancino}, {Pauwels}, {Petit}, {Recio-Blanco}, {Robin}, {Sarro},
  {Siopis}, {Smith}, {Smith}, {Sozzetti}, {Thuillot}, {van Reeven}, {Viala},
  {Abbas}, {Abreu Aramburu}, {Accart}, {Aguado}, {Allan}, {Allasia},
  {Altavilla}, {{\'A}lvarez}, {Alves}, {Anderson}, {Andrei}, {Anglada Varela},
  {Antiche}, {Antoja}, {Ant{\'o}n}, {Arcay}, {Atzei}, {Ayache}, {Bach},
  {Baker}, {Balaguer-N{\'u}{\~n}ez}, {Barache}, {Barata}, {Barbier}, {Barblan},
  {Baroni}, {Barrado y Navascu{\'e}s}, {Barros}, {Barstow}, {Becciani},
  {Bellazzini}, {Bellei}, {Bello Garc{\'\i}a}, {Belokurov}, {Bendjoya},
  {Berihuete}, {Bianchi}, {Bienaym{\'e}}, {Billebaud}, {Blagorodnova},
  {Blanco-Cuaresma}, {Boch}, {Bombrun}, {Borrachero}, {Bouquillon}, {Bourda},
  {Bouy}, {Bragaglia}, {Breddels}, {Brouillet}, {Br{\"u}semeister},
  {Bucciarelli}, {Budnik}, {Burgess}, {Burgon}, {Burlacu}, {Busonero}, {Buzzi},
  {Caffau}, {Cambras}, {Campbell}, {Cancelliere}, {Cantat-Gaudin}, {Carlucci},
  {Carrasco}, {Castellani}, {Charlot}, {Charnas}, {Charvet}, {Chassat},
  {Chiavassa}, {Clotet}, {Cocozza}, {Collins}, {Collins}, \& {Costigan}}]{gaia}
{Gaia Collaboration}, {Prusti}, T., {de Bruijne}, J.~H.~J., {et~al.} 2016,
  \aap, 595, A1

\bibitem[{{Guiglion} {et~al.}(2024){Guiglion}, {Nepal}, {Chiappini},
  {Khoperskov}, {Traven}, {Queiroz}, {Steinmetz}, {Valentini}, {Fournier},
  {Vallenari}, {Youakim}, {Bergemann}, {M{\'e}sz{\'a}ros}, {Lucatello},
  {Sordo}, {Fabbro}, {Minchev}, {Tautvai{\v{s}}ien{\.{e}}}, {Mikolaitis}, \&
  {Montalb{\'a}n}}]{rvs-cnn}
{Guiglion}, G., {Nepal}, S., {Chiappini}, C., {et~al.} 2024, \aap, 682, A9

\bibitem[{{Hattori}(2025)}]{hattori2025}
{Hattori}, K. 2025, \apj, 980, 90

\bibitem[{{Hunt} \& {Reffert}(2021)}]{hunt2021}
{Hunt}, E.~L. \& {Reffert}, S. 2021, \aap, 646, A104

\bibitem[{{Hunt} \& {Reffert}(2023)}]{hunt2023}
{Hunt}, E.~L. \& {Reffert}, S. 2023, \aap, 673, A114

\bibitem[{{Kane} {et~al.}(2025){Kane}, {Belokurov}, {Cranmer}, {Monty},
  {Zhang}, \& {Ardern-Arentsen}}]{kane2025}
{Kane}, S.~G., {Belokurov}, V., {Cranmer}, M., {et~al.} 2025, \mnras, 536, 2507

\bibitem[{{Khalatyan} {et~al.}(2024){Khalatyan}, {Anders}, {Chiappini},
  {Queiroz}, {Nepal}, {dal Ponte}, {Jordi}, {Guiglion}, {Valentini}, {Torralba
  Elipe}, {Steinmetz}, {Pantaleoni-Gonz{\'a}lez}, {Malhotra},
  {Jim{\'e}nez-Arranz}, {Enke}, {Casamiquela}, \&
  {Ard{\`e}vol}}]{khalatyan2024}
{Khalatyan}, A., {Anders}, F., {Chiappini}, C., {et~al.} 2024, \aap, 691, A98

\bibitem[{Kharchenko {et~al.}(2005)Kharchenko, Piskunov, Röser, Schilbach, \&
  Scholz}]{kharchenko2005}
Kharchenko, N.~V., Piskunov, A.~E., Röser, S., Schilbach, E., \& Scholz, R.-D.
  2005, \aap, 438, 1163

\bibitem[{{Kharchenko} {et~al.}(2013){Kharchenko}, {Piskunov}, {Schilbach},
  {R{\"o}ser}, \& {Scholz}}]{kharchenko2013}
{Kharchenko}, N.~V., {Piskunov}, A.~E., {Schilbach}, E., {R{\"o}ser}, S., \&
  {Scholz}, R.~D. 2013, \aap, 558, A53

\bibitem[{{Kroupa}(2002)}]{kroupa2002}
{Kroupa}, P. 2002, Science, 295, 82

\bibitem[{{Li} {et~al.}(2024){Li}, {Wong}, {Hogg}, {Rix}, \&
  {Chandra}}]{aspgap}
{Li}, J., {Wong}, K. W.~K., {Hogg}, D.~W., {Rix}, H.-W., \& {Chandra}, V. 2024,
  \apjs, 272, 2

\bibitem[{{Liu} \& {Pang}(2019)}]{liu2019}
{Liu}, L. \& {Pang}, X. 2019, \apjs, 245, 32

\bibitem[{{Lucey} {et~al.}(2023){Lucey}, {Al Kharusi}, {Hawkins}, {Ting},
  {Ramachandra}, {Price-Whelan}, {Beers}, {Lee}, \& {Yoon}}]{lucey2023}
{Lucey}, M., {Al Kharusi}, N., {Hawkins}, K., {et~al.} 2023, \mnras, 523, 4049

\bibitem[{{Magrini} {et~al.}(2023){Magrini}, {Viscasillas V{\'a}zquez},
  {Spina}, {Randich}, {Romano}, {Franciosini}, {Recio-Blanco}, {Nordlander},
  {D'Orazi}, {Baratella}, {Smiljanic}, {Dantas}, {Pasquini}, {Spitoni},
  {Casali}, {Van der Swaelmen}, {Bensby}, {Stonkute}, {Feltzing}, {Sacco},
  {Bragaglia}, {Pancino}, {Heiter}, {Biazzo}, {Gilmore}, {Bergemann},
  {Tautvai{\v{s}}ien{\.{e}}}, {Worley}, {Hourihane}, {Gonneau}, \&
  {Morbidelli}}]{gaia-eso-oc-mapping}
{Magrini}, L., {Viscasillas V{\'a}zquez}, C., {Spina}, L., {et~al.} 2023, \aap,
  669, A119

\bibitem[{{Marocco} {et~al.}(2021){Marocco}, {Eisenhardt}, {Fowler},
  {Kirkpatrick}, {Meisner}, {Schlafly}, {Stanford}, {Garcia}, {Caselden},
  {Cushing}, {Cutri}, {Faherty}, {Gelino}, {Gonzalez}, {Jarrett}, {Koontz},
  {Mainzer}, {Marchese}, {Mobasher}, {Schlegel}, {Stern}, {Teplitz}, \&
  {Wright}}]{catwise}
{Marocco}, F., {Eisenhardt}, P. R.~M., {Fowler}, J.~W., {et~al.} 2021, \apjs,
  253, 8

\bibitem[{{Martin} {et~al.}(2024){Martin}, {Starkenburg}, {Yuan}, {Fouesneau},
  {Ardern-Arentsen}, {De Angeli}, {Gran}, {Montelius}, {Rusterucci}, {Andrae},
  {Bellazzini}, {Montegriffo}, {Esselink}, {Zhang}, {Venn}, {Viswanathan},
  {Aguado}, {Battaglia}, {Bayer}, {Bonifacio}, {Caffau}, {C{\^o}t{\'e}},
  {Carlberg}, {Fabbro}, {Fern{\'a}ndez-Alvar}, {Gonz{\'a}lez Hern{\'a}ndez},
  {Gonz{\'a}lez Rivera de La Vernhe}, {Hill}, {Ibata}, {Jablonka},
  {Kordopatis}, {Lardo}, {McConnachie}, {Navarrete}, {Navarro}, {Recio-Blanco},
  {S{\'a}nchez-Janssen}, {Sestito}, {Thomas}, {Vitali}, \&
  {Youakim}}]{pristine-gaia}
{Martin}, N.~F., {Starkenburg}, E., {Yuan}, Z., {et~al.} 2024, \aap, 692, A115

\bibitem[{Monteiro {et~al.}(2019)Monteiro, Dias, Moitinho, \&
  Lépine}]{monteiro2019}
Monteiro, H., Dias, W.~S., Moitinho, A., \& Lépine, J. R.~D. 2019, \mnras,
  487, 2385

\bibitem[{{Netopil} {et~al.}(2016){Netopil}, {Paunzen}, {Heiter}, \&
  {Soubiran}}]{Netopil16}
{Netopil}, M., {Paunzen}, E., {Heiter}, U., \& {Soubiran}, C. 2016, \aap, 585,
  A150

\bibitem[{Palakkatharappil \& Creevey(2023)}]{Palakkatharappil23}
Palakkatharappil, D.~B. \& Creevey, O.~L. 2023, \aap, 674, A146

\bibitem[{{Perren} {et~al.}(2015){Perren}, {V\'azquez}, \& {Piatti}}]{asteca}
{Perren}, G.~I., {V\'azquez}, R.~A., \& {Piatti}, A.~E. 2015, A\&A, 576, A6

\bibitem[{{Randich} {et~al.}(2022){Randich}, {Gilmore}, {Magrini}, {Sacco},
  {Jackson}, {Jeffries}, {Worley}, {Hourihane}, {Gonneau}, {Viscasillas
  Vazquez}, {Franciosini}, {Lewis}, {Alfaro}, {Allende Prieto}, {Bensby},
  {Blomme}, {Bragaglia}, {Flaccomio}, {Fran{\c{c}}ois}, {Irwin}, {Koposov},
  {Korn}, {Lanzafame}, {Pancino}, {Recio-Blanco}, {Smiljanic}, {Van Eck},
  {Zwitter}, {Asplund}, {Bonifacio}, {Feltzing}, {Binney}, {Drew}, {Ferguson},
  {Micela}, {Negueruela}, {Prusti}, {Rix}, {Vallenari}, {Bayo}, {Bergemann},
  {Biazzo}, {Carraro}, {Casey}, {Damiani}, {Frasca}, {Heiter}, {Hill},
  {Jofr{\'e}}, {de Laverny}, {Lind}, {Marconi}, {Martayan}, {Masseron},
  {Monaco}, {Morbidelli}, {Prisinzano}, {Sbordone}, {Sousa}, {Zaggia},
  {Adibekyan}, {Bonito}, {Caffau}, {Daflon}, {Feuillet}, {Gebran}, {Gonzalez
  Hernandez}, {Guiglion}, {Herrero}, {Lobel}, {Maiz Apellaniz}, {Merle},
  {Mikolaitis}, {Montes}, {Morel}, {Soubiran}, {Spina}, {Tabernero},
  {Tautvai{\v{s}}iene}, {Traven}, {Valentini}, {Van der Swaelmen}, {Villanova},
  {Wright}, {Abbas}, {Aguirre B{\o}rsen-Koch}, {Alves}, {Balaguer-Nunez},
  {Barklem}, {Barrado}, {Berlanas}, {Binks}, {Bressan}, {Capuzzo-Dolcetta},
  {Casagrande}, {Casamiquela}, {Collins}, {D'Orazi}, {Dantas}, {Debattista},
  {Delgado-Mena}, {Di Marcantonio}, {Drazdauskas}, {Evans}, {Famaey},
  {Franchini}, {Fr{\'e}mat}, {Friel}, {Fu}, {Geisler}, {Gerhard}, {Gonzalez
  Solares}, {Grebel}, {Gutierrez Albarran}, {Hatzidimitriou}, {Held},
  {Jim{\'e}nez-Esteban}, {J{\"o}nsson}, {Jordi}, {Khachaturyants},
  {Kordopatis}, {Kos}, {Lagarde}, {Mahy}, {Mapelli}, {Marfil}, {Martell},
  {Messina}, {Miglio}, {Minchev}, {Moitinho}, {Montalban}, {Monteiro},
  {Morossi}, {Mowlavi}, {Mucciarelli}, {Murphy}, {Nardetto}, {Ortolani},
  {Paletou}, {Palou{\v{s}}}, {Paunzen}, {Pickering}, {Quirrenbach}, {Re
  Fiorentin}, {Read}, {Romano}, {Ryde}, {Sanna}, {Santos}, {Seabroke},
  {Spagna}, {Steinmetz}, {Stonkut{\'e}}, {Sutorius}, {Th{\'e}venin}, {Tosi},
  {Tsantaki}, {Vink}, {Wright}, {Wyse}, {Zoccali}, {Zorec}, {Zucker}, \&
  {Walton}}]{gaia-eso-randich}
{Randich}, S., {Gilmore}, G., {Magrini}, L., {et~al.} 2022, \aap, 666, A121

\bibitem[{{Riello} {et~al.}(2021){Riello}, {De Angeli}, {Evans}, {Montegriffo},
  {Carrasco}, {Busso}, {Palaversa}, {Burgess}, {Diener}, {Davidson}, {Rowell},
  {Fabricius}, {Jordi}, {Bellazzini}, {Pancino}, {Harrison}, {Cacciari}, {van
  Leeuwen}, {Hambly}, {Hodgkin}, {Osborne}, {Altavilla}, {Barstow}, {Brown},
  {Castellani}, {Cowell}, {De Luise}, {Gilmore}, {Giuffrida}, {Hidalgo},
  {Holland}, {Marinoni}, {Pagani}, {Piersimoni}, {Pulone}, {Ragaini}, {Rainer},
  {Richards}, {Sanna}, {Walton}, {Weiler}, \& {Yoldas}}]{riello2021}
{Riello}, M., {De Angeli}, F., {Evans}, D.~W., {et~al.} 2021, \aap, 649, A3

\bibitem[{{Rix} {et~al.}(2022){Rix}, {Chandra}, {Andrae}, {Price-Whelan},
  {Weinberg}, {Conroy}, {Fouesneau}, {Hogg}, {De Angeli}, {Naidu}, {Xiang}, \&
  {Ruz-Mieres}}]{rix2022}
{Rix}, H.-W., {Chandra}, V., {Andrae}, R., {et~al.} 2022, \apj, 941, 45

\bibitem[{{Sch{\"a}lte} {et~al.}(2022){Sch{\"a}lte}, {Klinger}, {Alamoudi}, \&
  {Hasenauer}}]{pyabc}
{Sch{\"a}lte}, Y., {Klinger}, E., {Alamoudi}, E., \& {Hasenauer}, J. 2022,
  JOSS, 7, 4304

\bibitem[{{Sim} {et~al.}(2019){Sim}, {Lee}, {Ann}, \& {Kim}}]{sim2019}
{Sim}, G., {Lee}, S.~H., {Ann}, H.~B., \& {Kim}, S. 2019, JKoAS, 52, 145

\bibitem[{{Spina} {et~al.}(2022){Spina}, {Magrini}, \& {Cunha}}]{spina2022}
{Spina}, L., {Magrini}, L., \& {Cunha}, K. 2022, Universe, 8, 87

\bibitem[{{Taylor}(2005)}]{topcat}
{Taylor}, M.~B. 2005, in Astronomical Society of the Pacific Conference Series,
  Vol. 347, Astronomical Data Analysis Software and Systems XIV, ed.
  P.~{Shopbell}, M.~{Britton}, \& R.~{Ebert}, 29

\bibitem[{{von Hippel} {et~al.}(2006){von Hippel}, {Jefferys}, {Scott},
  {Stein}, {Winget}, {De Gennaro}, {Dam}, \& {Jeffery}}]{base9}
{von Hippel}, T., {Jefferys}, W.~H., {Scott}, J., {et~al.} 2006, \apj, 645,
  1436

\bibitem[{{Yao} {et~al.}(2024){Yao}, {Ji}, {Koposov}, \& {Limberg}}]{yao2024}
{Yao}, Y., {Ji}, A.~P., {Koposov}, S.~E., \& {Limberg}, G. 2024, \mnras, 527,
  10937

\bibitem[{{Ye} {et~al.}(2025){Ye}, {Wu}, {Allende Prieto}, {Aguado}, {Zhao},
  {Gonz{\'a}lez Hern{\'a}ndez}, {Rebolo}, {Zhao}, {Li}, {del Burgo}, \&
  {Chen}}]{ye2025}
{Ye}, X., {Wu}, W., {Allende Prieto}, C., {et~al.} 2025, \aap, 695, A75

\bibitem[{{Ying} {et~al.}(2023){Ying}, {Chaboyer}, {Boudreaux}, {Slaughter},
  {Boylan-Kolchin}, \& {Weisz}}]{Ying2023}
{Ying}, J.~M., {Chaboyer}, B., {Boudreaux}, E.~M., {et~al.} 2023, \aj, 166, 18

\bibitem[{{Zhang} {et~al.}(2023){Zhang}, {Green}, \& {Rix}}]{zhang2023}
{Zhang}, X., {Green}, G.~M., \& {Rix}, H.-W. 2023, \mnras, 524, 1855

\end{thebibliography}
\appendix
\onecolumn
\section{[Fe/H] estimates}
\begin{figure*}[ht!]
	\centering
	\includegraphics[width=1\linewidth]{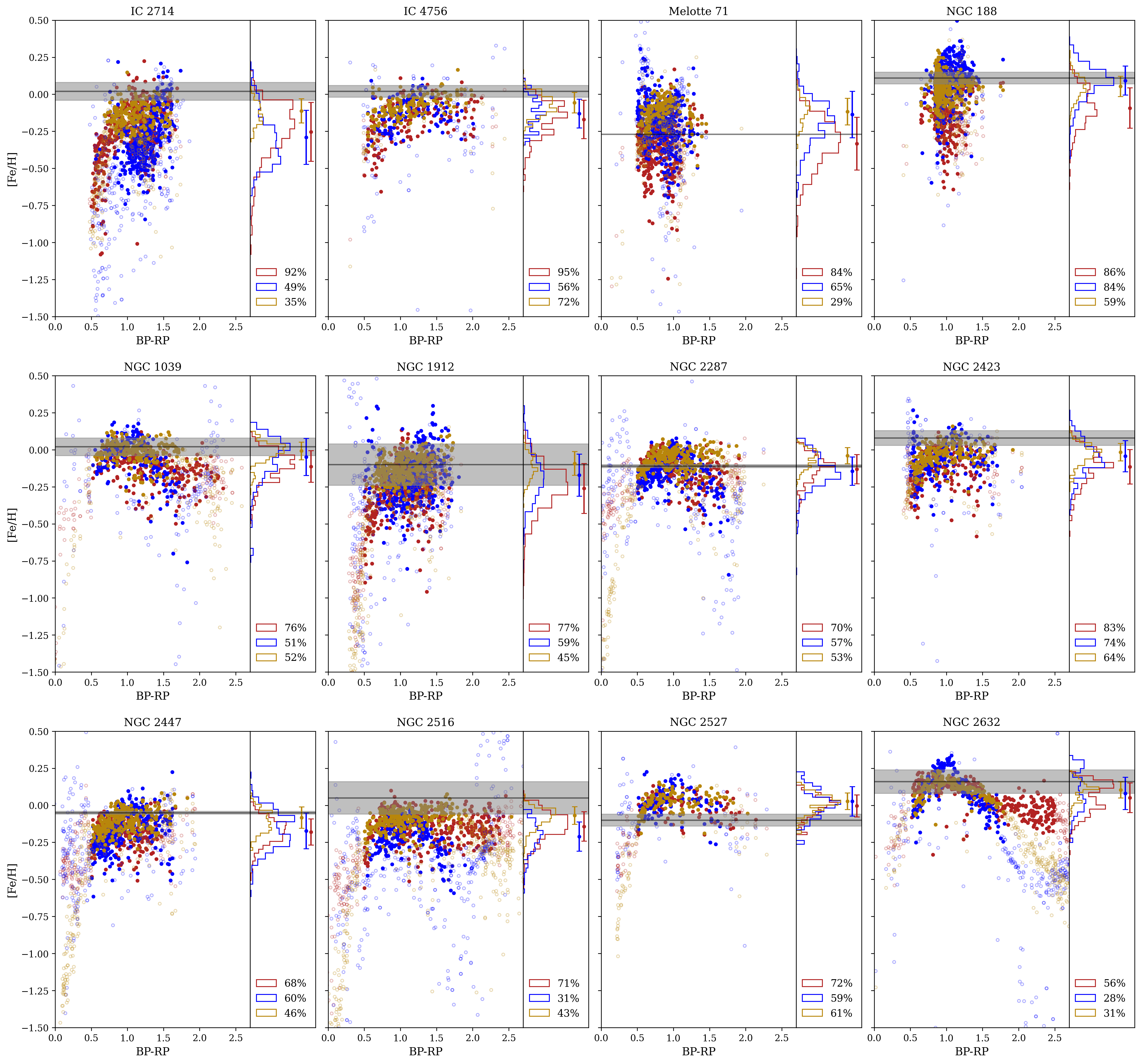}
	\caption{[Fe/H] estimates from three catalogues shown as a function of Gaia \bprp\ colour for each cluster of the sample. The same colour-coding scheme and legend description apply as in Fig. \ref{fig:bprp_met}. The black line and the shaded area represent the HRS [Fe/H] and $\sigma$[Fe/H] determined by \cite{Netopil16} and used by B19. The points with error bars show the weighted mean [Fe/H] and its dispersion of the respective catalogue, these values are listed in Table \ref{appendix_params_table1}. Melotte 71 does not depict the error of the HRS [Fe/H] measurement since \cite{Netopil16} obtained it from one star.}
	\label{fig:bprp_all_1}
\end{figure*}

\begin{figure*}[ht!]
	\centering
	\includegraphics[width=1\linewidth]{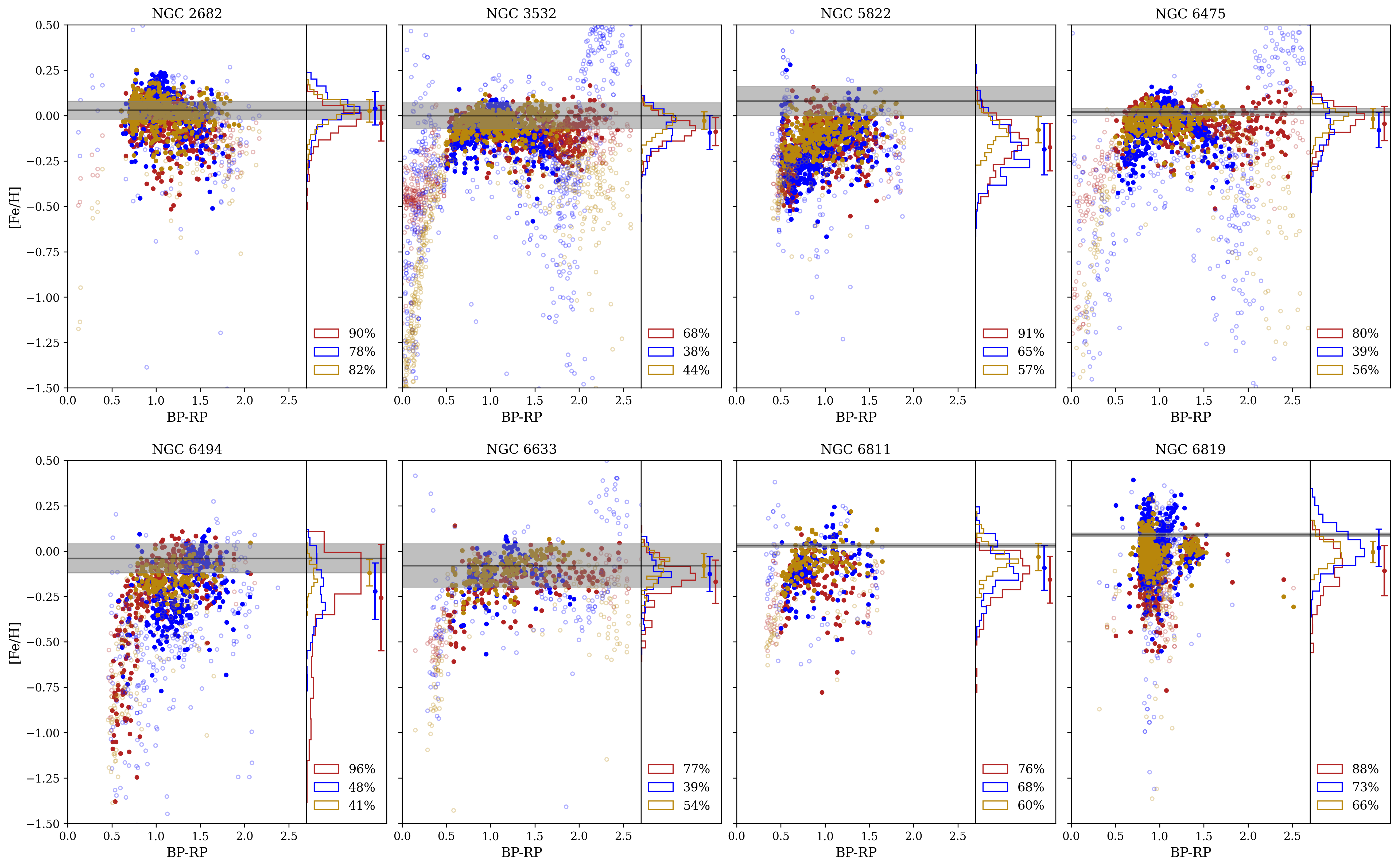}
	\caption{[Fe/H] estimates from three catalogues shown as a function of Gaia \bprp\ colour for each cluster of the sample. The same colour-coding scheme and legend description apply as in Fig. \ref{fig:bprp_all_1}.}
	\label{fig:bprp_all_2}
\end{figure*}

Figures \ref{fig:bprp_all_1} and \ref{fig:bprp_all_2} show the [Fe/H] estimates from the catalogues ARC, ZGR, and FS for individual stars of each cluster in our sample.
\newpage
\onecolumn
\section{Cluster parameters}
Parameters of 20 OCs derived with our analysis and the isochrones corresponding to the respective sets of parameters.
\begin{table*}[ht!]
	\centering
	\caption{List of clusters and their [Fe/H] estimates subsequently used as prior assumptions.} \label{appendix_params_table1}
	\begin{tabular}{llllll}
		\hline
		Cluster&RA&Dec&[Fe/H]*&[Fe/H] &$\sigma$ [Fe/H] \\
		\hline
		&(deg)&(deg)&&(dex)&(dex)\\
		\hline
		IC 2714    & 169.369 & -62.734 & ARC           & -0.254     & 0.199         \\
		&         &         & ZGR           & -0.291     & 0.183         \\
		&         &         & FS            & -0.112     & 0.081         \\
		&         &         & None          & 0.256      & -             \\
		&         &         & HRS           & 0.02       & -             \\
		IC 4756    & 279.627 & 5.437   & ARC           & -0.17      & 0.129         \\
		&         &         & ZGR           & -0.13      & 0.097         \\
		&         &         & FS            & -0.057     & 0.069         \\
		&         &         & None          & 0.374      & -             \\
		&         &         & HRS           & 0          & -             \\
		Melotte 71 & 114.377 & -12.058 & ARC           & -0.334     & 0.178         \\
		&         &         & ZGR           & -0.137     & 0.157         \\
		&         &         & FS            & -0.117     & 0.09          \\
		&         &         & None          & 0.093      & -             \\
		&         &         & HRS           & -0.27      & -             \\
		...         & ...     & ...     & ...           & ...        & ...          
	\end{tabular}
		\tablefoot{[Fe/H]* indicates if the metallicity was derived from a Gaia XP-based catalogue (ARC, ZGR, FS), or adopted as a HRS value from \citet{bossini2019}. ‘None’ indicates that [Fe/H] was a free fit parameter, and the 50th percentile of its PDF is listed. The entire table is available at the CDS.}
\end{table*}
\begin{table*}[ht!]
	\centering
	\caption{Cluster parameters A$_V$, dm and \logage\ derived using the specified [Fe/H]* prior assumption.} \label{appendix_params_table2}
	\begin{tabular}{lllllllllll}
		\hline 
		Cluster&[Fe/H]*&A$_V$&A$_{V16}$&A$_{V84}$&dm&dm$_{16}$&dm$_{84}$&log age&log age$_{16}$&log age$_{84}$\\
		\hline
		&&(mag)&(mag)&(mag)&(mag)&(mag)&(mag)&(dex)&(dex)&(dex)\\
		\hline
		IC 2714    & ARC           & 1.182 & 1.146  & 1.23   & 10.31  & 10.218 & 10.368 & 8.678  & 8.581      & 8.707      \\
		& ZGR           & 1.211 & 1.148  & 1.253  & 10.307 & 10.238 & 10.341 & 8.661  & 8.448      & 8.720       \\
		& FS            & 1.188 & 1.145  & 1.232  & 10.324 & 10.286 & 10.396 & 8.674  & 8.473      & 8.704      \\
		& None          & 1.075 & 0.983  & 1.191  & 10.577 & 10.475 & 10.656 & 8.671  & 7.859      & 8.729      \\
		& HRS           & 1.029 & 1.008  & 1.15   & 10.318 & 10.298 & 10.375 & 8.657  & 8.656      & 8.659      \\
		IC 4756    & ARC           & 0.953 & 0.716  & 1.1    & 8.267  & 8.152  & 8.43   & 8.770   & 8.766      & 9.033      \\
		& ZGR           & 0.784 & 0.667  & 0.996  & 8.2    & 8.135  & 8.295  & 9.031  & 8.767      & 9.034      \\
		& FS            & 0.617 & 0.559  & 0.648  & 8.226  & 8.194  & 8.254  & 9.013  & 8.966      & 9.014      \\
		& None          & 0.396 & 0.302  & 0.472  & 8.441  & 8.332  & 8.511  & 8.946  & 8.917      & 8.994      \\
		& HRS           & 0.566 & 0.539  & 0.611  & 8.231  & 8.209  & 8.249  & 9.003  & 9.001      & 9.004      \\
		Melotte 71 & ARC           & 0.514 & 0.471  & 0.55   & 11.486 & 11.404 & 11.529 & 9.131  & 9.107      & 9.152      \\
		& ZGR           & 0.468 & 0.415  & 0.496  & 11.576 & 11.533 & 11.609 & 9.088  & 9.069      & 9.099      \\
		& FS            & 0.468 & 0.448  & 0.493  & 11.568 & 11.528 & 11.604 & 9.088  & 9.075      & 9.094      \\
		& None          & 0.437 & 0.326  & 0.484  & 11.702 & 11.56  & 11.82  & 8.994  & 8.967      & 9.083      \\
		& HRS           & 0.537 & 0.523  & 0.547  & 11.472 & 11.45  & 11.539 & 9.138  & 9.136      & 9.141      \\
		...         & ...           & ...   & ...    & ...    & ...    & ...    & ...    & ...    & ...        & ...       
	\end{tabular}
	\tablefoot{The [Fe/H]* notation is defined as in Table \ref{appendix_params_table1}. The 50th, 16th and 84th percentile of the PDF for each cluster parameter is reported. The entire table is available at the CDS.}
\end{table*}
\begin{figure}[ht!]
	\centering
	\includegraphics[width=1\linewidth]{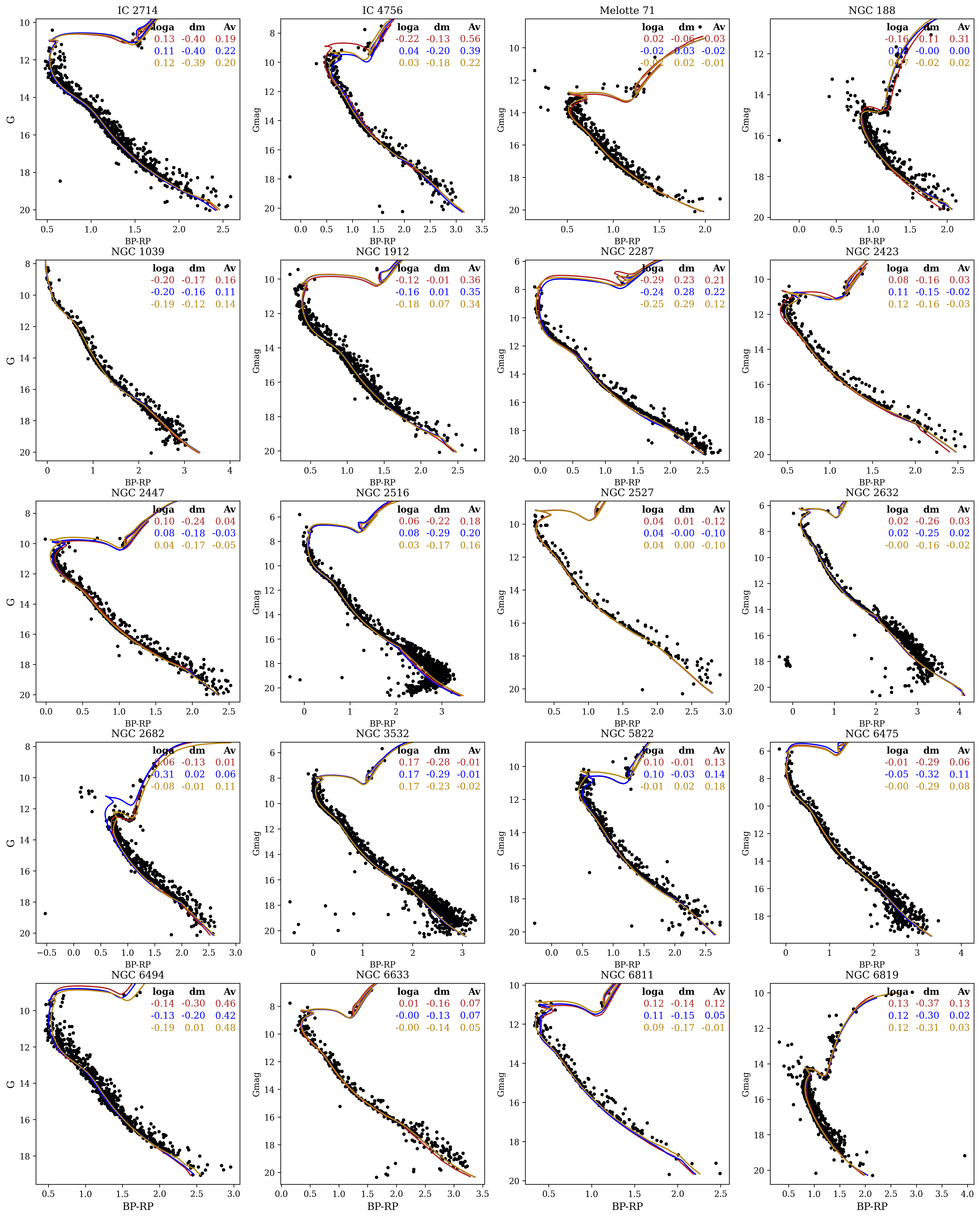}
	\caption{CMDs of the sample of OCs with photometric data from Gaia. The isochrones match the cluster parameters derived by using [Fe/H] of ZGR (blue), ARC (red), and FS (yellow) catalogues; the depicted values show the difference between the parameters (columns) estimated by using the respective prior and B19 values of the same parameter. }
	\label{fig:all-cmds}
\end{figure}
\begin{figure*}[ht]
	\centering
	\begin{subfigure}[b]{0.48\textwidth}
		\centering
		\includegraphics[width=\textwidth]{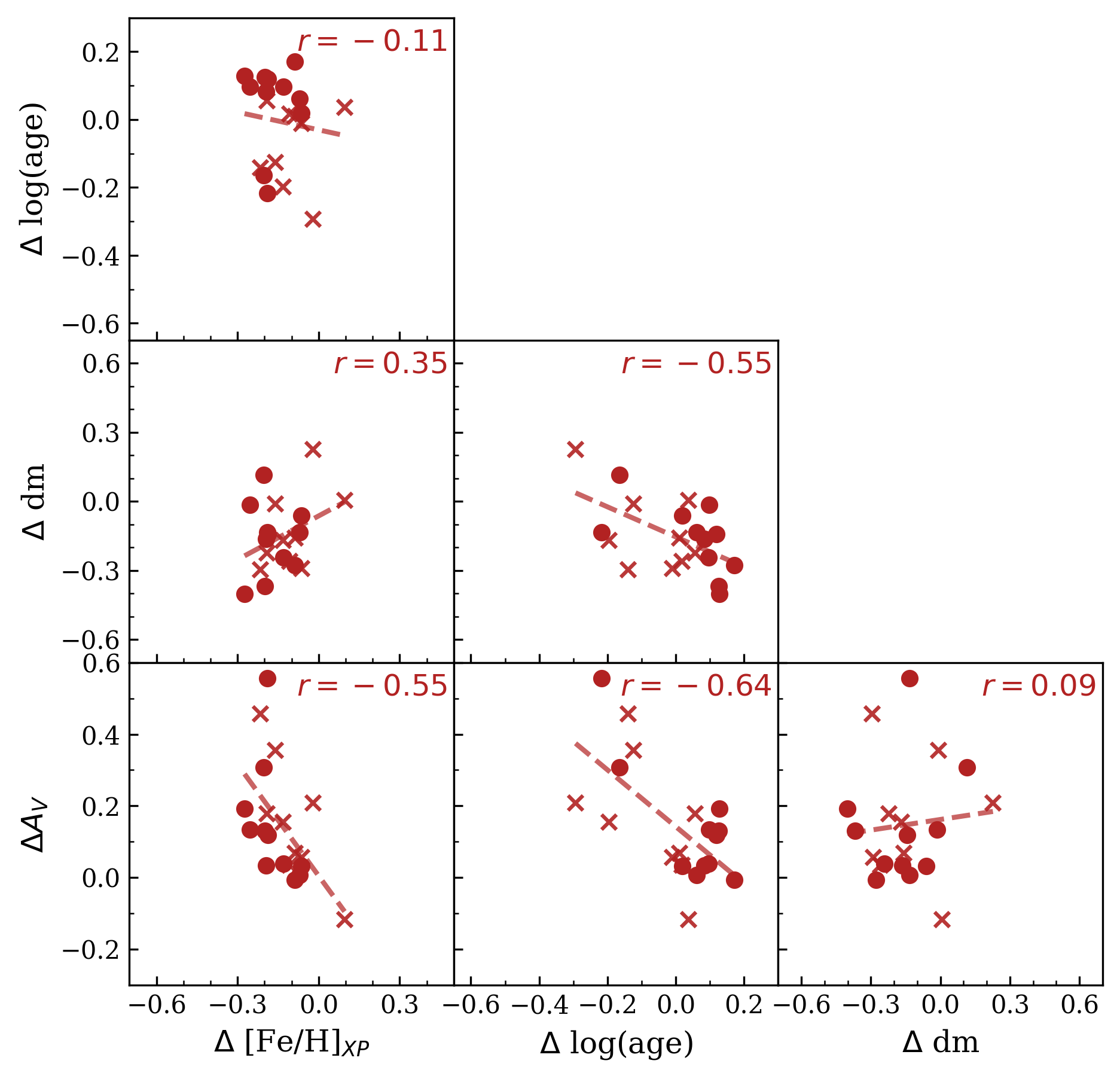}
		\caption{}
		\label{fig:corner_arc}
	\end{subfigure}
	\hfill
	\begin{subfigure}[b]{0.48\textwidth}
		\centering
		\includegraphics[width=\textwidth]{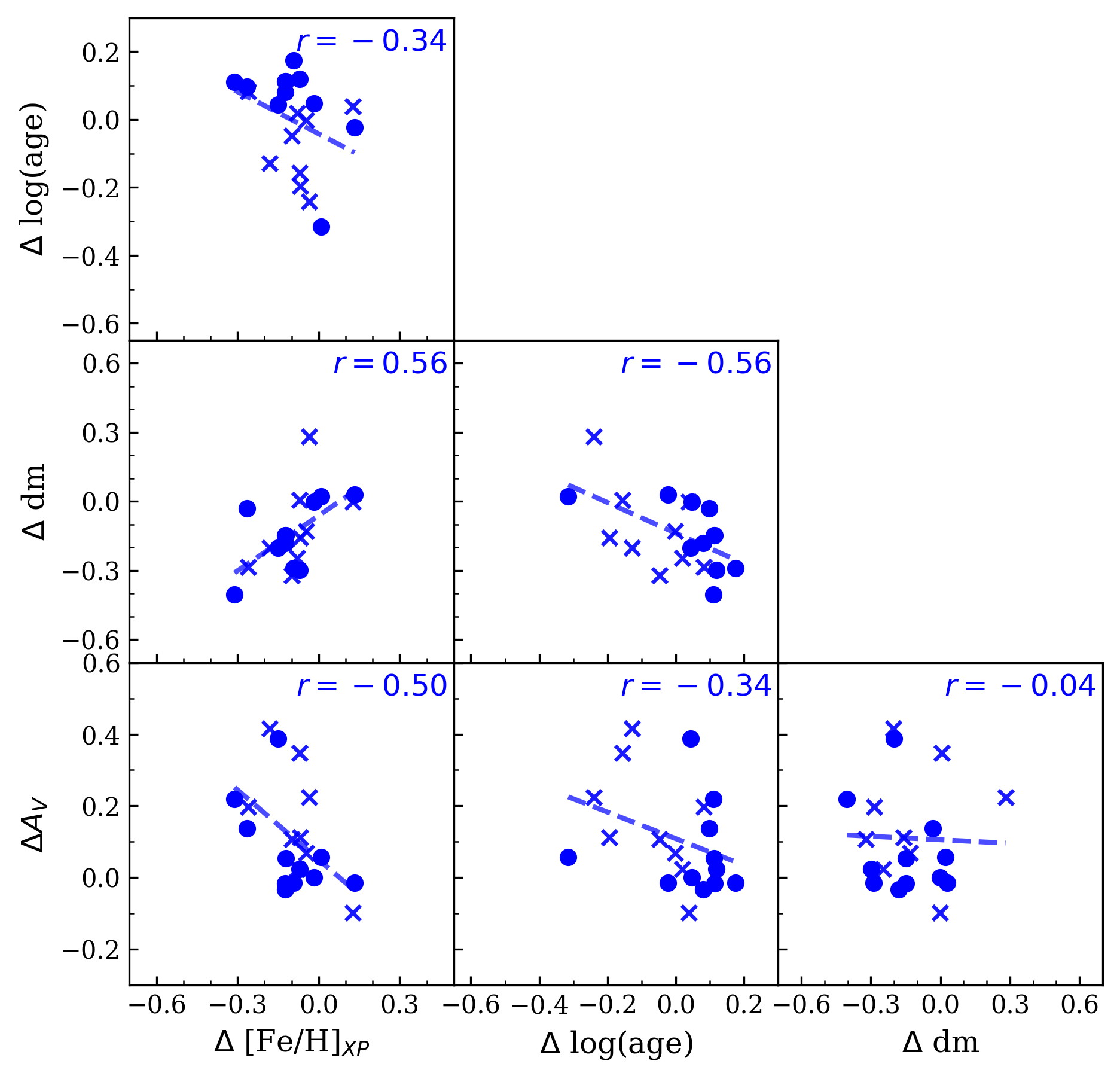}
		\caption{}
		\label{fig:corner_zgr}
	\end{subfigure}
	\caption{Same as Figure \ref{fig:corner_nofix_fs} but for (a) ARC catalogue and (b) ZGR catalogue.}
	\label{fig:corner_arc_zgr}
\end{figure*}
\begin{figure}[ht]
	\centering
	\includegraphics[width=0.45\textwidth]{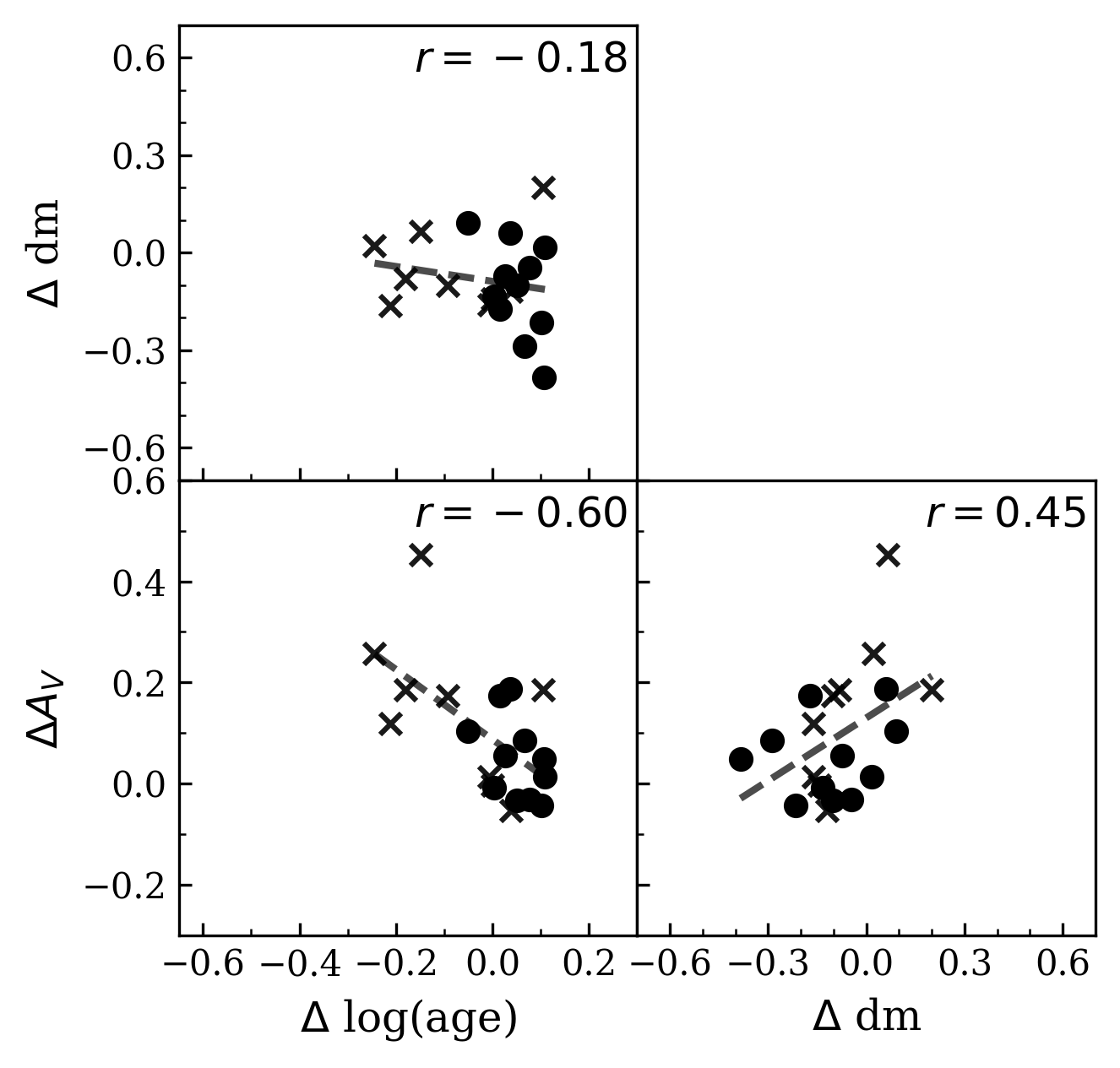}
	\caption{Same as Figure \ref{fig:corner_nofix_fs} but for solutions with [Fe/H] fixed to the HRS values from B19. The [Fe/H] offset is not displayed, as it is zero.}
	\label{fig:corner_fix}
\end{figure}
\end{document}